\documentclass{jfm}
\usepackage{graphicx}
\usepackage{epstopdf, epsfig}
\usepackage{newtxtext}
\usepackage{newtxmath}

\usepackage{hyperref}
\usepackage{wrapfig}
\usepackage{amsfonts}
\usepackage[ruled,vlined]{algorithm2e}
\newsavebox{\tempfig}
\usepackage[utf8]{inputenc}
\usepackage[english]{babel}
\usepackage{caption}
\usepackage{amsmath}
\usepackage{amssymb}
\usepackage{amsfonts} 
\usepackage{mathrsfs}
\usepackage{mathtools}
\usepackage{physics}
\usepackage{xfrac}
\usepackage[utf8]{inputenc}
\usepackage[english]{babel}
\usepackage[table,usenames,dvipsnames]{xcolor}
 \usepackage{textcomp}
 \usepackage[table]{xcolor}
 \usepackage{multirow}
 \usepackage{mathrsfs,amsmath}
\usepackage[square,numbers]{natbib}
\usepackage{arydshln}
\usepackage[makeroom]{cancel}
\usepackage{tikz}
\usetikzlibrary{plotmarks}
\usepackage{etoolbox}
\newcommand{\be}{\begin{equation}}
\newcommand{\ee}{\end{equation}}
\usepackage{graphicx}
\usepackage{etoolbox}
\usepackage{comment}
\DeclareRobustCommand\full  {\tikz[baseline=-0.6ex]\draw[blue,thick] (0,0)--(0.5,0);}
\DeclareRobustCommand\fullgreen{\tikz[baseline=-0.6ex]\draw[green!60!black,thick] (0,0)--(0.5,0);}
\DeclareRobustCommand\fullblack  {\tikz[baseline=-0.6ex]\draw[black,thick] (0,0)--(0.5,0);}
\DeclareRobustCommand\fullred  {\tikz[baseline=-0.6ex]\draw[red,thick] (0,0)--(0.5,0);}

\DeclareRobustCommand\dotted{\tikz[baseline=-0.6ex]\draw[black,thick,dotted] (0,0)--(0.54,0);}
\DeclareRobustCommand\dashedgreen{\tikz[baseline=-0.6ex]\draw[green!60!black,thick,dashed] (0,0)--(0.54,0);}
\DeclareRobustCommand\dottedgray{\tikz[baseline=-0.6ex]\draw[white!60!black,thick,dotted] (0,0)--(0.54,0);}

\DeclareRobustCommand\dashedblack{\tikz[baseline=-0.6ex]\draw[black,thick,dashed] (0,0)--(0.54,0);}
\DeclareRobustCommand\dashedgray{\tikz[baseline=-0.6ex]\draw[white!60!black,thick,dashed] (0,0)--(0.54,0);}

\DeclareRobustCommand\chainblack {\tikz[baseline=-0.6ex]\draw[black, thick,dash dot ] (0,0)--(0.5,0);}
\DeclareRobustCommand\chaingreen {\tikz[baseline=-0.6ex]\draw[green!60!black, thick,dash dot ] (0,0)--(0.5,0);}

\newrobustcmd*{\mycircle}[1]{\tikz{\filldraw[draw=#1,fill=#1] (0,0) circle [radius=0.05cm];}}
\newrobustcmd*{\mytriangle}[1]{\tikz{\filldraw[draw=#1,fill=#1] (0,0)--(0.2cm,0) -- (0.1cm,0.2cm);}}
\newrobustcmd*{\mysquare}[1]{\tikz{\filldraw[draw=#1,fill=#1] (0,0)--(0.15cm,0) -- (0.15cm,0.15cm) -- (0,0.15cm);}}

\def\del#1{\textcolor{gray}{}}

\shorttitle{Vorticity-drag relation in bluff body flows}
\shortauthor{Y. Du and T. A. Zaki}

\title{Vorticity dynamics and drag for flows over a sphere and a prolate spheroid}

\author{Yifan Du\aff{1}
 \and Tamer A. Zaki\aff{1}\corresp{\email{t.zaki@jhu.edu}}}

\affiliation{\aff{1}Department of Mechanical Engineering, Johns Hopkins University,
Baltimore, MD 21218, USA}

\begin{document}

\maketitle

\begin{abstract}
The connection between the drag and vorticity dynamics for viscous flow over a bluff body is explored using the Josephson-Anderson (J-A) relation for classical fluids. The instantaneous rate of work on the fluid, associated with the drag force, is related to the vorticity flux across the streamlines of a background potential flow. The vorticity transport itself is examined by aid of the Huggins vorticity flux tensor. The analysis is performed for three flows: flow over a sphere at Reynolds numbers $\Rey=\{200,3700\}$ and flow over a prolate spheroid at $\Rey=3000$ and $20^{\circ}$ incidence. In these flows, the vorticity transport shifts the flow away and towards the ideal potential flow, with a net balance towards the former effect thus making an appreciably contribution to the drag. The J-A relation is first demonstrated for the flow over a sphere at $\Rey=200$.  The drag power injection is related to the viscous flux of azimuthal vorticity from the wall into the fluid and the advection of vorticity by the detached shear layer. In the wake, the azimuthal vorticity is advected towards the wake centerline and is annihilated by viscous effects, which contributes a reduction to drag. The analysis of the flow over a sphere at $\Rey=3700$ is reported for the impulsively started and stationary stages, with emphasis on the effects of unsteady two-dimensional separation and turbulent transport in the transitional wake. The turbulent flux in the wake enhances the transport of mean azimuthal vorticity towards the wake centerline, and is the main driver of the recovery of enthalpy between the rear point of the sphere and far downstream. 
The rate of work on the fluid by the drag force for a prolate spheroid is mostly due to the transport of vorticity along the separated boundary layers. Primary and secondary separation contribute oppositely to the power injection by the drag force, while the large-scale vortices only re-distribute vorticity without a net contribution. A mechanism for secondary separation is proposed based on the theory of vortex-induced separation.
\end{abstract}

\vspace*{12pt}


\section{Introduction}

The relationship between drag and the transport of vorticity is established for some canonical flows, perhaps most known among them are the flows in channels and pipes \citep{taylor1932transport}. In these configurations, the transverse transport of vorticity leads to a drag force in the streamwise direction. Such connection provides a special perspective to interpret drag in viscous, vorticity- and vortex-dominated flows.  
For flows over immersed bodies, this connection is given by the detailed Josephson-Anderson relation that equates the rate of work done by the drag force and the flux of vorticity against a background potential flow \citet{eyink2021josephson}. 
In this study, we examine this relation for three-dimensional separated flows over spheres at moderate Reynolds number and over a spheroid at incidence.   

The description of vorticity transport involves the notion of vorticity flux.  In the conservative form of the vorticity equation, the Huggins flux tensor compactly captures the vorticity evolution by advection, tilting and stretching, and viscous diffusion \citep{huggins1970energy, huggins1971dynamical}. 
However, the vorticity equation only dictates the form of the vorticity-flux tensor up to a divergence-free contribution.  A different form of the vorticity flux was introduced by \citet{Lighthill} and \citet{panton1984incompressible} at the solid wall, and subsequently extended into the fluid interior \citep{kolar2003lyman}. 
The physical interpretations of Huggins and Lighthill-Panton flux tensors were discussed and compared by \citet{terrington2021generation}. Both definitions similarly capture the spatial transport of vorticity, although the two forms contain different viscous contributions. The Huggins flux tensor measures the viscous transfer of circulation due to tangential acceleration in the fluid, while the viscous part of Lighthill-Panton flux only includes the terms that lead to the local change of vorticity. 
In the present study, the Huggins' definition is adopted because it is uniquely related to the momentum and force balances. The turbulent part of the vorticity flux, which corresponds to Reynolds stress in the momentum equations, has been used to characterize the structure of near-wall turbulence \citep{klewicki2013description}. Recently, the Huggins flux tensor has been extensively applied to examine the vorticity transport in flow over rotating and translating spheres \citep{terrington2021generation} and free-surface flows \citep{terrington2022vortex, terrington2022vorticity}.  Here, we will quantify the vorticity motion for the flows over a sphere and a spheroid.

The connection between drag and vorticity fluxes has been the subject of influential works.  \citet{taylor1932transport} introduced the correspondence between streamwise pressure gradient and the transverse transport of spanwise vorticity in shear flows. \citet{Lighthill} quantified the generation of vorticity at solid walls and established the balance between wall vorticity-flux and the pressure gradient in channel and pipe flows.  For the particular flows of interest in this work, namely flows over isolated bluff bodies, the drag force has been expressed as the integral of physical quantities within the flow interior. For example, \citet{lighthill1986informal} indirectly related the rate of work done by the drag force with the viscous dissipation of kinetic energy. Through a similar argument, \citet{stone1993interpretation} equated the rate of work done by the drag force over a steadily translating drop with the energy dissipation in the surrounding fluid and within the drop itself. The total dissipation is further decomposed into contributions from enstrophy and an interface vorticity term associated with surface curvature. \citet{howe1995force} and \citet{magnaudet2011reciprocal} expressed the drag force as an integral formula in terms of velocity and vorticity fields. 
In particular, \citet{howe1995force} constructed potential flows using different boundary conditions, and expressed lift and torque using these potentials. 
The work by \citet{eyink2021josephson}, although derived through a different approach, 
yielded a similar expression that he termed the detailed Josephson-Anderson (J-A) relation. 
This expression relates the drag force to vorticity transport, and thus provides fluid dynamical insight. These ideas were demonstrated recently for the interpretation of drag in unsteady, separated laminar flow over a hill \citep{kumar2024JA}.  Here our focus is on bluff-body flows. Using the J-A relation, we describe the rate of work done on the fluid by the drag force in terms of the vorticity flux crossing the potential-flow streamlines.  This connection highlights important regions of the flow. In addition, the J-A relation can be interpreted in terms of an energy transfer between the vortical flow and the ideal, or potential, flow around the object.  These ideas provide new perspectives on the role of vorticity and its dynamics in bluff-body flows and the generation of drag, which we explore in the present work for the flows around a sphere and a prolate spheroid.

The flow around a sphere exhibits a wealth of fluid dynamical phenomena, which have been the subject of experimental and numerical studies.  The early works by \citet{achenbach1974vortex} and \citet{sakamoto1990study} systematically documented the drag force and the flow states over a wide range of Reynolds numbers, $400\leq\Rey\leq 5\times10^{6}$ based on the sphere diameter. 
Direct numerical simulations (DNS) and large eddy simulations (LES) were performed in the sub-critical range $\Rey<\Rey_c=3.7\times10^{5}$, where separation is laminar \citep{johnson1999flow, constantinescu2004numerical, yun2006vortical,  rodriguez2011direct,bazilevs2014computation}. The work by \citet{johnson1999flow} documented several flow regimes, including steady axisymmetric flow ($\Rey\leq 200$), steady non-axisymmetric flow ($210\leq\Rey\leq270$), and planar vortex shedding ($\Rey=300$). Based on DNS at $\Rey=3700$, \citet{rodriguez2011direct} and \citet{bazilevs2014computation} reported the flow statistics, morphology of the vortices, and the shedding mechanism. Vortices form and shed at random azimuthal angles due to the change in wall pressure, which results in a helical-shape wake. \citet{yun2006vortical} performed LES at $\Rey=\{3700, 10^{4}\}$ and found that the higher $\Rey$ case exhibits a smaller recirculation region, and earlier transition and recovery of the wake. These studies establish a qualitative physical picture about the motion of vortices using visualization of vortical structures, passive particle tracers, and frequency analysis of the shedding process.  We will complement those efforts by providing a quantitative analysis, based on the Huggins flux tensor and directly linking the recovery of the wake to the vorticity transport. 

Compared to the sub-critical flow over a sphere, additional complexity is introduced when three-dimensional separation develops over the surface of a bluff body. An important example is the flow over a spheroid at incidence.  In an early study, \citet{wang1990three} systematically investigated the friction lines and separation patterns on prolate spheroids with different aspect ratios and incidence angles, at subcritical Reynolds number.  They discovered the phenomenon of open separation, and showed that multiple separations and re-attachments can appear on the leeward side depending on the incidence angle. 
\citet{ahn1992experimental} and \citet{wetzel1996unsteady} identified the critical Reynolds number, $\Rey_c\sim 4.2\times 10^{5}$ based on the length of the minor axis of a $6\colon1$ prolate spheroid, where transition occurs in the boundary layer prior to separation.
They documented the flow statistics and friction patterns at both the sub- and super-critical conditions. \citet{fu1994flow} studied the counter-rotating vortices and the crossflow separations above the leeward surface, and related the circulation in the large-scale vortices to the lift and lateral forces. Such relation underscores the importance of vorticity dynamics in the near-body field.  Direct numerical simulations of flow over a spheroid have been limited to the sub-critical flow regimes \citep{el2010crossflow,jiang2016peculiar}, and have focused primarily on the vortical structures in the turbulent wake rather than of the separated boundary layer. 
\citet{jiang2016peculiar} reported a non-symmetric helical wake behind a $6\colon1$ prolate spheroid at $45^{\circ}$ incidence, and identified the axial separation line as the origin of the vortex generation.  \citet{ortiz2021high} performed LES at $\Rey=10^{5}$ based on the length of the spheroid minor axis, and proposed a new decay rate for the velocity deficit based on the non-equilibrium dissipation scaling in the far wake. More recent LES by \citet{Plasseraud2023} demonstrated that tripping the boundary layer has a limited effect at $\Rey=7\times10^{5}$ and $20^{\circ}$ incidence. 
For a theoretical treatment, \citet{wu2000vorticity} and \citet{surana2006exact} introduced criteria for the identification of three-dimensional separation, and discussed the relationship between vorticity and separation. 
Also relevant to the present effort are studies of vortex-induced separation \citep{peridier1991vortex,doligalski1994vortex}, since the dominant counter-rotating vortices on the leeward side of the spheroid can interact with the underlying boundary layer.   Building of these previous efforts, we will directly evaluate the motion of axial vorticity using the flux tensor, and provide a vorticity-based mechanism for separation on the spheroid.

In section \ref{sec:JA}, we start by introducing the theoretical framework that is adopted for our study, including the setup of the flow over a bluff body, the governing equations, and the J-A relation. The numerical simulations of the flows over a sphere and a prolate spheroid are presented in section \ref{sec:numerical simulation}. The J-A relation and the Huggins flux tensor are numerically evaluated.  We first present the analysis of laminar flow over a sphere at $\Rey=200$ as a preliminary example in section  \ref{subsection:Re200}. The more complex case of an impulsively started flow and the late-stage stationary state over the sphere at $\Rey=3700$ are analyzed in sections \ref{subsection:Re3700 starting} and \ref{subsection:Re3700}, with a focus on the two dimensional unsteady separation and the turbulent wake dynamics.  We then proceed to discuss the vorticity transport in the three-dimensional separation on the prolate spheroid in section \ref{sec:spheroid results}. A summary and conclusion are provided in section \ref{sec:conclusions}

\section{Theoretical formulation and computational approach}
\label{sec:JA}

In this section, we introduce the computational setup for simulating the flow over bluff bodies, including the domain geometries, the governing equations, and the boundary conditions.  We discuss the vorticity dynamics, including the Helmholtz equation and the Huggins flux tensor. We then provide a brief derivation of the Josephson-Anderson relation, which is the theoretical approach that we adopt to interpret the power injection into the fluid in terms of vorticity fluxes. 


\subsection{Flow over bluff body}\label{subsec:Flow over bluff body}

\begin{figure}
    \centering
    \includegraphics[width=1.0\textwidth]{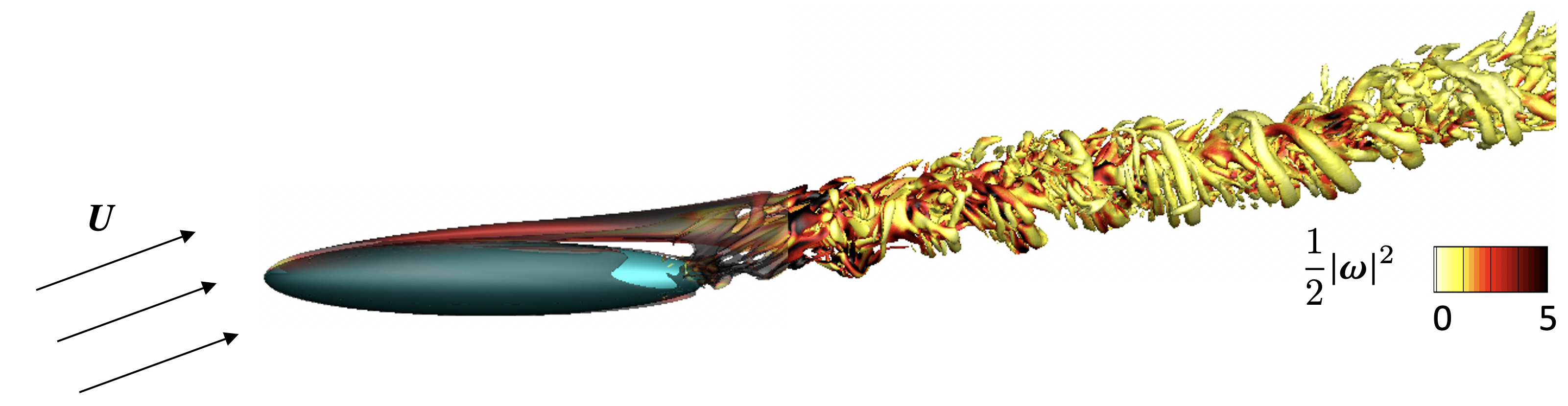}
    \caption{Schematic of the flow over a bluff body. A uniform flow with velocity $\boldsymbol{U}$ passes over a solid body $B$ (coloured in blue). The vortical structures are defined by the iso-surface of the Q-criteria $Q=0.5$, and are coloured by enstrophy.}
    \label{fig:flow over bluff body}
\end{figure}

We consider the flow over a bluff body, which is depicted schematically in figure \ref{fig:flow over bluff body}. A solid isolated body occupying a spatial volume $B$ is held stationary. The space outside $B$ is filled with a viscous incompressible fluid with viscosity $\nu^{*}$, where the superscript star designates dimensional quantities. The non-dimensional form of the governing Navier-Stokes equations is: 
\begin{eqnarray}\label{eq:NS}
    \label{eq:div}
    \nabla \cdot \boldsymbol{u} &=& 0,   \\
	\label{eq:mom}
	\frac{\partial \boldsymbol{u} }{\partial t} &=& \boldsymbol{u} \times \boldsymbol{\omega}  - \nabla h + \frac{1}{\Rey} \nabla^2 \boldsymbol{u},
\end{eqnarray}
where $\boldsymbol{u}=(u,v,w)$ is the velocity vector, and the generalized enthalpy $h\equiv \frac{p}{\rho}+\frac{1}{2}|\boldsymbol{u}|^2$ is the sum of the pressure (divided by density which is normalized to unity) and kinetic energy per unit mass. The bulk Reynolds number is $\Rey=\left|\boldsymbol{U}^{*}\right|L^{*}/\nu^{*}$, where $\boldsymbol{U}^{*}$ is the free-stream velocity. The characteristic length $L^{*}$ is defined in terms of the size of the bluff body, using the diameter of the sphere and the length of the minor axis for the spheroid. Since the bluff body is stationary, the velocity field at $\partial B$ satisfies the no-slip boundary conditions. In the far field $|\boldsymbol{x}| \rightarrow \infty$, the flow approaches the free-stream velocity. These boundary conditions are expressed as: 
\begin{equation}
    \left.\boldsymbol{u}\right|_{\partial B}=0, \qquad \boldsymbol{u} \underset{|\boldsymbol{x}| \rightarrow \infty}{\sim} \boldsymbol{U}, 
\end{equation}
where $\boldsymbol{U} = \boldsymbol{U}^{*}/\left|\boldsymbol{U}^{*}\right|$ is the non-dimensional free-stream velocity. The body exerts a drag force on the fluid, which is expressed as the surface integral of pressure and wall shear stress, 
\begin{equation}\label{eq:drag force}\boldsymbol{F} = \int_{\partial B}\left(p\hat{\boldsymbol{n}}-2\mu\boldsymbol{S}\hat{\boldsymbol{n}}\right)\mathrm{d}S, \qquad\boldsymbol{S}=\frac{1}{2}\left(\nabla\boldsymbol{u}+\nabla\boldsymbol{u}^{\top}\right), 
\end{equation}
where $\hat{\boldsymbol{n}}$ is the unit normal vector on the surface of the body pointing into the fluid, $\mu$ is the dynamic viscosity, and $\boldsymbol{S}$ is the strain rate tensor. The definition of $\boldsymbol{F}$ is the force exerted by the body on the fluid, consistent with \citet{eyink2021josephson}. We adopt this convention because the momentum and energy balances considered in the derivation of the Josephson-Anderson relation are for the fluid domain. The governing equations (\ref{eq:NS}) and initial and boundary conditions provide a complete description of the velocity field and of the drag force. 

Taking the curl of the momentum equation (\ref{eq:mom}) yields the Helmholtz equation for the vorticity $\boldsymbol{\omega}\equiv\nabla\times\boldsymbol{u}$, 
\begin{equation}\label{eq:Helmholtz}
    \frac{\partial \boldsymbol{\omega}}{\partial t} + \boldsymbol{\nabla} \cdot \boldsymbol{\Sigma} = 0, 
\end{equation}
where 
\begin{equation}\label{eq:Huggins tensor}
    \boldsymbol{\Sigma} = \boldsymbol{u}\boldsymbol{\omega}-\boldsymbol{\omega}\boldsymbol{u}-\nu \left(\nabla \boldsymbol{\omega} - \nabla\boldsymbol{\omega}^{\top}\right)
\end{equation}
is the Huggins flux tensor. The Helmholtz equation (\ref{eq:Helmholtz}) is written as a conservation law by aid of the Huggins flux tensor $\boldsymbol{\Sigma}$ which encodes the transport of vorticity by advection, stretching and tilting, and viscous diffusion. Specifically, the entries of $\Sigma_{ij}$ represent the flux of the $j$-th vorticity component in the $i$-th coordinate direction. The Huggins tensor is anti-symmetric, and thus admits a dual representation by an axial vector $\boldsymbol{\eta}$, 
\begin{equation}\label{eq:axial representation}
    \boldsymbol{\eta}=\boldsymbol{u}\times\boldsymbol{\omega}-\nu\nabla\times\boldsymbol{\omega}, \qquad \eta_i = \frac{1}{2}\varepsilon_{ijk}\Sigma_{jk},\qquad\Sigma_{ij}=\varepsilon_{ijk}\eta_k. 
\end{equation}
Note that the vorticity-flux tensor in equation (\ref{eq:Helmholtz}) is not uniquely defined. Any tensor $\boldsymbol{\Sigma}' = \boldsymbol{\Sigma} + \boldsymbol{\Theta}$ with $\nabla \cdot \boldsymbol{\Theta}=0$ is a valid option that describes the same vorticity evolution as $\boldsymbol{\Sigma}$. However, the Huggins flux tensor $\boldsymbol{\Sigma}$ is the unique choice that can be connected with momentum transport and pressure gradient. This connection is given by the rotational form of momentum equation, 
\begin{equation}\label{eq:mom_S}
    \frac{\partial u_i}{\partial t}=\frac{1}{2}\epsilon_{ijk}\Sigma_{jk}-\frac{\partial h}{\partial x_i}.
\end{equation}
Equation (\ref{eq:mom_S}) provides a direct relation between the vorticity flux and the total-pressure gradient, which at stationary walls reduces to,
\begin{equation}\label{eq:mom_S wall}
    -\nu\nabla\times\boldsymbol{\omega}=\nabla p. 
\end{equation}
An expression for the wall-vorticity flux $\boldsymbol{\sigma}=\boldsymbol{n}\cdot\boldsymbol{\Sigma}$ follows by taking the cross product between the wall-normal unit-vector and equation (\ref{eq:mom_S wall}), 
\begin{equation}\label{eq: wall vorticity flux}
\boldsymbol{\sigma}=\boldsymbol{n}\cdot\boldsymbol{\Sigma}=\boldsymbol{n}\times\left(\nu\nabla\times\boldsymbol{\omega}\right)=-\boldsymbol{n}\times\nabla p,  
\end{equation}
where $\boldsymbol{n}= -\hat{\boldsymbol{n}}$ is the unit normal vector on the wall pointing towards the body. The physical meaning of $\boldsymbol{\sigma}$ is the rate of vorticity transport from the fluid out through the wall per unit surface area. Integrating the vorticity equation (\ref{eq:Helmholtz}) in a finite domain enclosed by a surface, the time rate of change of vorticity in this domain is balanced by the surface integral of $\boldsymbol{\sigma}$. According to equation (\ref{eq: wall vorticity flux}), the viscous vorticity flux at the wall is closely related with the tangential pressure gradient, as discussed by \citet{Lighthill} and \citet{morton1984generation}. In this study, we numerically evaluate the Huggins vorticity flux (\ref{eq:Huggins tensor}) and use it to examine the transport of vorticity. We also explore the role of vorticity flux in the momentum balance using equation (\ref{eq:mom_S}). 

We end this section with a brief discussion of empirical observations related to boundary-layer separation and vorticity. The boundary layer develops along the surface of the body from the attachment line and evolves downstream, and the curvature of the surface induces favorable and adverse pressure gradients. Since the Reynolds numbers in the numerical studies considered in this work are subcritical ($\Rey_c=3\times 10^{5}$ for flow over sphere \citep{achenbach1974vortex} and $4.2\times 10^{5}$ for flow over spheroid \citep{ahn1992experimental}), natural transition  to turbulence does not take place prior to separation. Furthermore, depending on the geometry and inflow conditions, the boundary layer undergoes either two- or three-dimensional separation as can be gleaned from the wall shear stress $\boldsymbol{\tau}_w=2\mu\boldsymbol{S}\hat{\boldsymbol{n}}$ \citep{tobak1982topology}.  For instance, the steady axisymmetric flow over a sphere at $24\leq\Rey\left(:= \frac{U^{*}D^{*}}{\nu^{*}}\right)\leq 200$ undergoes two-dimensional separation at a polar angle $0^{\circ}<\theta<62^{\circ}$ where $\tau_w=0$ \citep{johnson1999flow}. For an example of three-dimensional separation, consider flow over a prolate spheroid with a non-zero angle of attack \citep{wang1990three}. Depending on the angle of incidence, multiple separation and re-attachment patterns can develop and result in a complex topology of wall-stress lines on the spheroid surface.  Separation lines are identified as limiting friction lines connecting two singular points of the wall shear stress (where $\tau_w=0$) and have neighboring wall shear stress lines converging towards them \citep{chapman1991topology,surana2006exact}. 

In both 2-D and 3-D separation, the motion of vorticity is key to understanding the boundary-layer dynamics. First, the vorticity is directly related to the wall shear stress by $\boldsymbol{\tau}_w=\mu\boldsymbol{\omega}\times\hat{\boldsymbol{n}}$, thus the characterization of separation using the wall shear stress can be equivalently established in terms of the vorticity \citep{wu2000vorticity}. Secondly, the wall pressure gradient is related to the vorticity flux through equation (\ref{eq: wall vorticity flux}) \citep{wu2007vorticity}. Favourable pressure gradient creates vorticity with the same orientation as that within the boundary layer, which keeps the boundary layer attached to the wall; Adverse pressure gradient creates vorticity with opposite orientation to the existing vorticity within the boundary layer, and thus promotes flow separation from the wall.


\subsection{The detailed Josephson-Anderson relation}
\label{subsec:Detailed Josephson-Anderson relation}

We start by stating the Josephson-Anderson (J-A) relation for flow over a stationary isolated body, and then proceed to summarize its derivation.  For the interested reader, a detailed derivation is provided by \citet{eyink2021josephson}. 
In the lab frame, the power injected into the fluids by the drag force is given by $-\boldsymbol{F}\cdot\boldsymbol{U}$. This quantity, in the body frame, can be expressed as,
\begin{eqnarray}\label{eq:JA}
    -\boldsymbol{F} \cdot \boldsymbol{U} &=& -\rho \int_{\Omega}\boldsymbol{u}_{\phi}\cdot\left(\boldsymbol{u}\times\boldsymbol{\omega}-\nu\boldsymbol{\nabla}\times\boldsymbol{\omega}\right)\mathrm{d}V\\ 
    \label{eq:JA1}
    &=& -\int\mathrm{d}J\int\left(\boldsymbol{u}\times\boldsymbol{\omega}-\nu\boldsymbol{\nabla}\times\boldsymbol{\omega}\right)\cdot\mathrm{d}\boldsymbol{l}\\ \label{eq:JA2}
    &=& -\frac{1}{2}\int\mathrm{d}J\int \varepsilon_{ijk}\Sigma _{ij}\mathrm{d}l_k. \label{eq:JA3}
\end{eqnarray}
This power injection is equal to the integral over the fluid volume $\Omega$ of the vorticity flux against a background potential flow $\boldsymbol{u}_{\phi}$, as shown by expression (\ref{eq:JA}). 
The volume element is decomposed into $\mathrm{d}V = \mathrm{d}A\left|\mathrm{d}\boldsymbol{l}\right|$, where $\mathrm{d}\boldsymbol{l}$ is a vector line element along $\boldsymbol{u}_{\phi}$, and $\mathrm{d}A$ is an area element normal to $\boldsymbol{u}_{\phi}$. By defining $\mathrm{d}J = \rho \left|\boldsymbol{u}_{\phi}\right|\mathrm{d}A$ as the potential mass current within a streamtube, the right-hand-side of the J-A relation is cast into expression (\ref{eq:JA1}).  Using the identities in (\ref{eq:axial representation}), the J-A relation is then written in terms of the Huggins flux tensor $\boldsymbol{\Sigma}$ in equation (\ref{eq:JA2}). We consider this right-hand side starting with $\mathrm{d}l_k$ which is the line element aligned with the potential flow, and notice that $\varepsilon_{ijk}\Sigma _{ij}\mathrm{d}l_k$ vanishes if either $i$ or $j$ are equal to $k$.  The implication is that contributions to the integral only arise due to components in the Huggins tensor that correspond to the flux ($i$-index) and the vorticity ($j$-index) both being orthogonal to the potential flow, that is when $i \ne k$ and $j \ne k$. Hence the expressions (\ref{eq:JA}-\ref{eq:JA2}) represent the the rate of energy injection associated with the vorticity flux crossing the potential-flow streamlines, weighted by the potential flow speed.  In other words, the rate of drag work performed by the immersed body onto the fluid is equal to the amount of vorticity normal to the potential flow that crosses the potential mass current outward. 
Conversely, inward vorticity flux across the potential-flow streamlines reduces drag power.  Lastly, vorticity flux along the potential-flow streamlines does not contribute to rate of work by drag.

The derivation of the J-A relation starts by defining $(\boldsymbol{u}_{\phi}, p_{\phi})$ as the potential-flow solution over the solid body with the same free-stream configuration, which can be obtained either analytically or numerically. The true velocity and pressure fields from the Navier-Stokes solution are then split into potential and vortical parts, with the vortical solution defined as: 
\begin{equation}
    \boldsymbol{u}_{\omega} \coloneqq \boldsymbol{u}-\boldsymbol{u}_{\phi}, \qquad p_{\omega} \coloneqq p-p_{\phi}. 
\end{equation}
The governing equations for the vortical flow $(\boldsymbol{u}_{\omega}, p_{\omega})$ is derived by subtracting the Euler and Navier-Stokes equations, which yields,
\begin{equation}\label{eq: vort mom}
    \partial_t \boldsymbol{u}_\omega=\boldsymbol{u} \times \boldsymbol{\omega}-\nu \boldsymbol{\nabla} \times \boldsymbol{\omega}-\boldsymbol{\nabla} h_\omega, 
\end{equation}
where $h_\omega=p_\omega+\frac{1}{2}\left|\boldsymbol{u}_\omega\right|^2+\boldsymbol{u}_\omega \cdot \boldsymbol{u}_\phi$ is the total pressure for the vortical flow. A relation between the total vortical momentum $\boldsymbol{P}_{\omega}$, the total vortical drag $\boldsymbol{F}_{\omega}$, and the far-field vortical pressure is obtained by integrating equation (\ref{eq: vort mom}) in space,
\begin{equation}\label{eq:vortical momentum--drag force}
    \frac{\mathrm{d} \boldsymbol{P}_\omega}{\mathrm{d} t}=\boldsymbol{F}_\omega-\rho \lim _{R \rightarrow \infty} \int_{S_R} \hat{\boldsymbol{x}} h_\omega \mathrm{d} S, \qquad \boldsymbol{P}_{\omega} =\int_{\Omega}\rho\boldsymbol{u}_{\omega}\mathrm{d}V ,\qquad \boldsymbol{F}_{\omega} = \int_{\partial B}\left(p_{\omega}\hat{\boldsymbol{n}}-2\mu\boldsymbol{S}\hat{\boldsymbol{n}}\right)\mathrm{d}S, 
\end{equation}
where $S_R$ is a sphere with radius $R$ and $\hat{\boldsymbol{x}} = \boldsymbol{x}/|\boldsymbol{x}|$ is the radially outward unit-vector. The total vortical momentum $\boldsymbol{P}_{\omega}$ represents the amount of momentum contained in the vortical flow. The total vortical force $\boldsymbol{F}_{\omega}$ is the same as the drag in equation (\ref{eq:drag force}). 
The first expression of \ref{eq:vortical momentum--drag force} shows that the rate of change of the total vortical momentum is driven by the imbalance between the drag force $\boldsymbol{F}_{\omega}$ and the far-field vortical pressure.  This time derivative ${d\boldsymbol{P}_\omega}/{dt}$ is generally not zero. 
In addition to the momentum equation (\ref{eq:vortical momentum--drag force}), we also consider the energy equation. The following expression for the local kinetic energy is obtained from the potential-vortical decomposition of velocity: 
\begin{equation}\label{eq: three energies}
    e = \frac{1}{2}\rho\boldsymbol{u}\cdot\boldsymbol{u} = \frac{1}{2}\rho\boldsymbol{u}_{\phi}\cdot\boldsymbol{u}_{\phi}+\rho\boldsymbol{u}_{\phi}\cdot\boldsymbol{u}_{\omega}+\frac{1}{2}\rho\boldsymbol{u}_{\omega}\cdot\boldsymbol{u}_{\omega}. 
\end{equation}
An energy evolution equation can then be written for each of the three ingredients, namely the potential, interaction and vortical kinetic energies. The volume-integral of the potential kinetic energy is infinite and conserved. We are specifically interested in the evolution equation of the interaction energy $\boldsymbol{u}_{\phi} \cdot \boldsymbol{u}_{\omega}$.  Performing a dot product of the governing equation for $\boldsymbol{u}_{\omega}$ with $\boldsymbol{u}_{\phi}$, we obtain
\begin{equation}\label{eq:interaction energy}
    \begin{aligned}
        \partial_{t}\left(\boldsymbol{u}_{\phi} \cdot \boldsymbol{u}_{\omega}\right)+\boldsymbol{\nabla} \cdot\left[h_{\omega} \boldsymbol{u}_{\phi}+h_\phi \boldsymbol{u}_{\omega}\right]=\boldsymbol{u}_{\phi} \cdot(\boldsymbol{u} \times \omega-\nu \boldsymbol{\nabla} \times \boldsymbol{\omega}), 
    \end{aligned}
\end{equation}
where $h_\phi=p_\phi+\frac{1}{2}\left|\boldsymbol{u}_\phi\right|^2$ is the potential total pressure. 
The divergence term spatially transports the local interaction energy, and does not cause any net change in the total interaction energy. The right-hand side is the work done by the vortex force $(\boldsymbol{u} \times \boldsymbol{\omega}-\nu \boldsymbol{\nabla} \times \boldsymbol{\omega})$ along the potential flow $\boldsymbol{u}_{\phi}$. 
Similar to momentum, equation (\ref{eq:interaction energy}) is integrated over the fluid domain,
\begin{equation}\label{eq:integrated interaction energy}
    \frac{\mathrm{d}}{\mathrm{d} t}\int_{\Omega}\rho\boldsymbol{u}_{\phi} \cdot \boldsymbol{u}_{\omega}\mathrm{d}V=+\rho \int_{\Omega} \boldsymbol{u}_\phi \cdot(\boldsymbol{u} \times \boldsymbol{\omega}-\nu \boldsymbol{\nabla} \times \boldsymbol{\omega}) d V-\rho \boldsymbol{U} \cdot \lim _{R \rightarrow \infty} \int_{S_R} \hat{\boldsymbol{x}} h_\omega d A.
\end{equation}
Additionally, a multi-pole expansion of the vortical velocity field $\boldsymbol{u}_{\omega}$ provides the following expression for the total interaction energy,
\begin{equation}\label{eq:interaction energy--vortical momentum}
    \int_{\Omega}\boldsymbol{u}_{\phi} \cdot \boldsymbol{u}_{\omega}\mathrm{d}V =  \int_{\Omega}\boldsymbol{u}_{\omega}\mathrm{d}V\cdot\boldsymbol{U}. 
\end{equation}
Combining equations (\ref{eq:vortical momentum--drag force}), (\ref{eq:integrated interaction energy}) and (\ref{eq:interaction energy--vortical momentum}) yields the J-A relation (\ref{eq:JA}), which is repeated below: 
\begin{equation}
        -\boldsymbol{F} \cdot \boldsymbol{U} = \rho \int_{\Omega}\underbrace{\boldsymbol{u}_{\phi}\cdot\left(-\boldsymbol{u}\times\boldsymbol{\omega}\right)}_{\Pi_a}+\underbrace{\boldsymbol{u}_{\phi}\cdot\left(\nu\boldsymbol{\nabla}\times\boldsymbol{\omega}\right)}_{\Pi_\nu}\mathrm{d}V. \tag{2.11}
\end{equation}

This relation introduces a new perspective on the transport of vorticity, the transfer of energy, and the drag power: First, the fields $\Pi_{a}=-\boldsymbol{u}_{\phi}\cdot\left(\boldsymbol{u}\times\boldsymbol{\omega}\right)$ and $\Pi_{\nu}=\boldsymbol{u}_{\phi}\cdot\left(\nu\boldsymbol{\nabla}\times\boldsymbol{\omega}\right)$ are the energy fluxes that correspond to advective and viscous vorticity transport crossing the potential streamlines of $\boldsymbol{u}_{\phi}$. Positive values represent advection and diffusion effects transporting negative vorticity outwards across potential streamlines, and the reverse for negative values. 
Secondly, $\Pi_{a}$ and $\Pi_{\nu}$ can be interpreted as spatial contributions to the power injection into the fluid by the drag force, with positive and negative values being the drag and anti-drag contributions. Finally, comparing the right-hand side of the J-A relation (\ref{eq:JA}) to the interaction-energy equation (\ref{eq:interaction energy}) shows that drag is associated with a loss in the interaction energy. This loss appears as a source in the vortical energy equation, 
\begin{equation}\label{eq:vortical energy}
    \begin{aligned}
      \partial_{t}\left(\frac{1}{2}\left|\boldsymbol{u}_{\omega}\right|^{2}\right)+\boldsymbol{\nabla} \cdot\left[\left(p_{\omega}+\frac{1}{2}\left|\boldsymbol{u}_{\omega}\right|^{2}+\boldsymbol{u}_{\omega} \cdot \boldsymbol{u}_{\phi}\right) \boldsymbol{u}_{\omega}-\nu \boldsymbol{u} \times \omega\right] \\
    =-\boldsymbol{u}_{\phi} \cdot(\boldsymbol{u} \times \boldsymbol{\omega}-\nu \boldsymbol{\nabla} \times \boldsymbol{\omega})-\nu|\boldsymbol{\omega}|^{2}.   
    \end{aligned}
\end{equation}
In other words,  $\Pi_a + \Pi_{\nu} = -\boldsymbol{u}_{\phi} \cdot(\boldsymbol{u} \times \boldsymbol{\omega}-\nu \boldsymbol{\nabla} \times \boldsymbol{\omega})$ is exactly equal to the energy transfer from the interaction energy to the vortical energy. Positive values are transfers from interaction to vortical energy by advection and viscous effects, and the reverse for negative values. Ultimately the vortical energy is dissipated into heat by the last term -$\nu|\boldsymbol{\omega}|^2$ in equation (\ref{eq:vortical energy}).

The physical interpretation of the J-A relation in a finite integration domain demands more careful discussion. 
A steady or statistically stationary state of the flow near the body can be reached at a sufficiently long time after an initial impulsive start. However, the flow far downstream remains non-stationary, for example in the region where the starting vortex is first observed.
As such, the integral of the unsteady term in equation (\ref{eq:interaction energy}) does not have a contribution from the stationary flow near the body, and is solely due to the far wake.  In other words, the rate of change of total interaction energy is primarily due to the far wake where the flow has not reached stationarity.  
The streamwise extent of this region of the wake is at least the product of the free-stream velocity and the transient time (from the initial condition to the stationary state), which is too large to include in a direct numerical simulation (DNS). 
Nevertheless, the Josephson-Anderson relation can still be approximately satisfied within the domains customarily adopted for DNS because the right-hand side of equation (\ref{eq:interaction energy}) decays fast downstream\textemdash an argument that is numerically verified in this study.

\subsection{Numerical simulation of flow over bluff body}\label{sec:numerical simulation}

\begin{figure}
    \centering
    \includegraphics[width=0.95\textwidth]{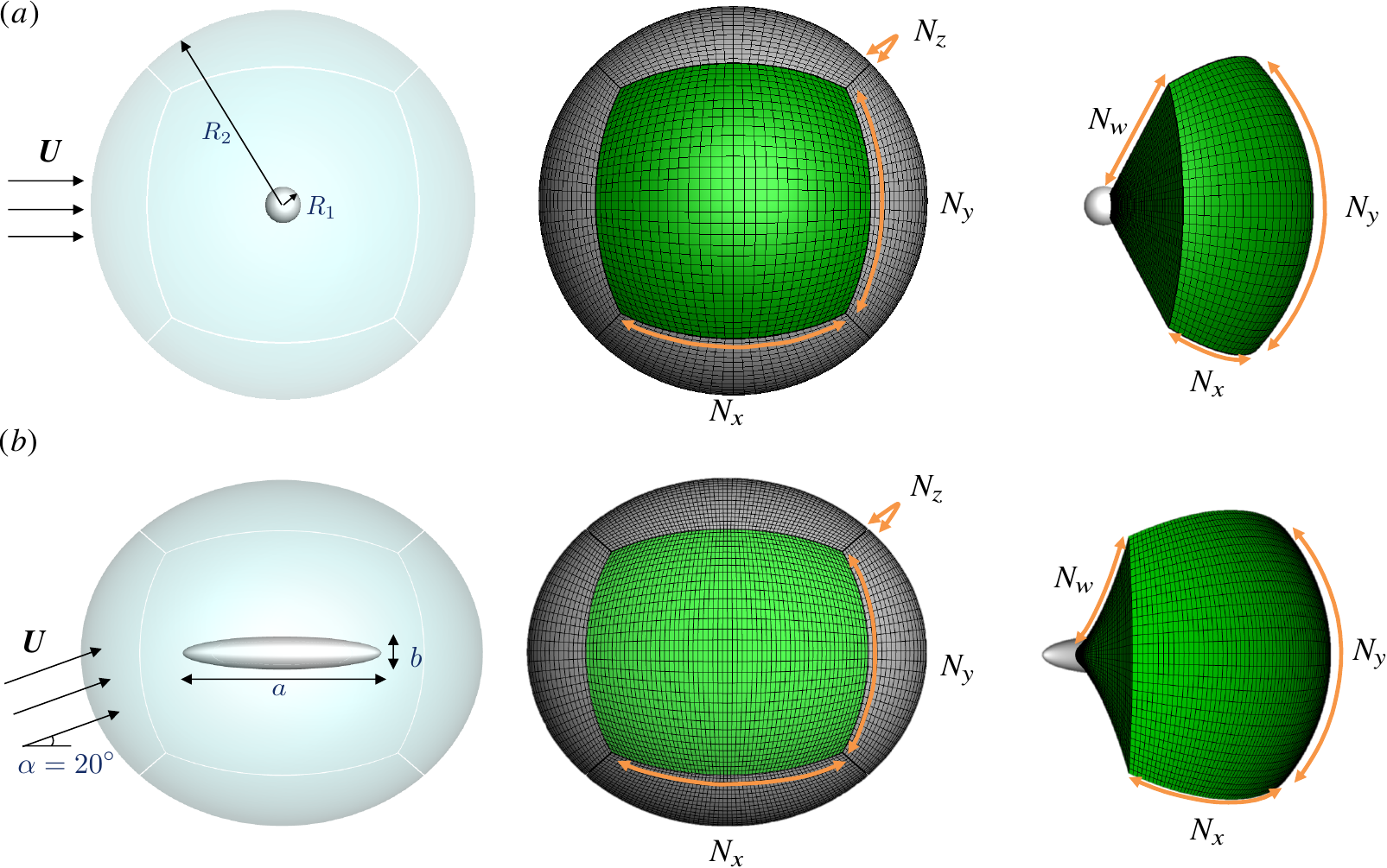}
    \caption{Computational domains and meshes. $(a)$ The flow setup, multi-block grid system, and a rotated view of the front block for the flow over the sphere. The same visualizations are repeated in $(b)$ for the flow over the prolate spheroid.}
    \label{fig: domain grid}
\end{figure}
\begin{figure}
    \centering
    \includegraphics[width=1.0\textwidth]{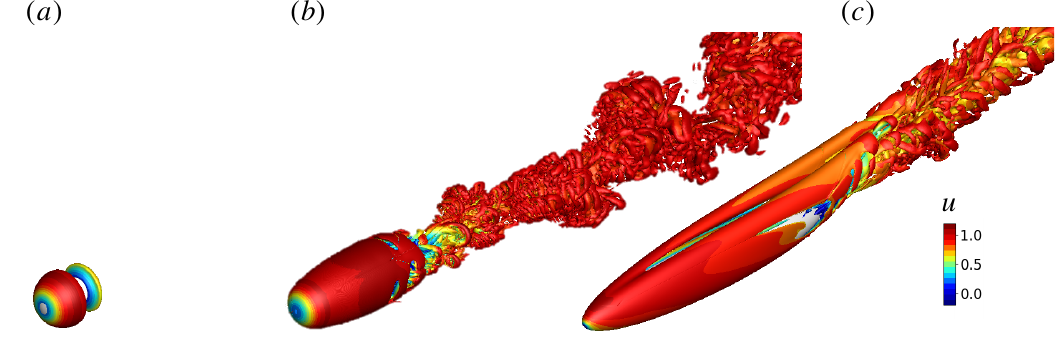}
    \caption{Vortical structures visualized using iso-surface of the Q-criteria and coloured by the streamwise velocity, from simulations of the flows over the spheres and the spheroid. $(a)$ Flow over a sphere at $\Rey=200$, $Q=0.5$.  $(b)$ Flow over a sphere at $\Rey=3700$, $Q=0.1$. $(c)$ Flow over a spheroid at $\Rey=3000$, $Q=0.5$. }
    \label{fig:vortical structures}
\end{figure}
\begin{table}
		\centering
		\begin{tabular}{c c c c c}
			Case & Geometry  & \ Reynolds number \  & Grid points & Grid resolution\\
                \begin{tabular}{c}
                \hline
                 \\ \hline 
				  SL \\ 
                    ST\\
				PT         \\
			\end{tabular} &
			\begin{tabular}{c}
                \hline
                Bluff body shape \\ \hline 
				  \multirow{2}{*}{Sphere} \\ 
                     \\
				Spheroid         \\
			\end{tabular} &
                \begin{tabular}{c}
				\hline
				$\Rey$ \\ \hline 
			      200                \\
        		3700                 \\
                    3000                 \\
			\end{tabular}&
                \begin{tabular}{c c c c}
				\hline
				$N_x$ & $N_y$ & $N_z$ & $N_w$ \\ \hline 
					
				129             &   129           &   129                     &  193  \\
    			257             &   161           &   161                     &  769  \\
        		241             &   241           &   241                     &  385  \\
			\end{tabular} &
			\begin{tabular}{c c}
				\hline
				$\Delta y_b$ & $\Delta_w$\\ \hline 
			      0.006  & 0.088         \\
        		0.001  & 0.013         \\
                    0.0017 & 0.028         \\
			\end{tabular}\\
		\end{tabular}
		\caption{Geometries, Reynolds numbers, number of grid points, and resolutions for direct numerical simulations. The resolution \(\Delta y_{b}\) represents the wall-normal grid spacing at the solid wall, while \(\Delta_w = \left(\Delta x_w \Delta y_w \Delta z_w\right)^{1/3}\) denotes the grid size at a point in the wake three units of length downstream of the trailing edge of the sphere and spheroid.}
		\label{table:numerical setup}
	\end{table}
In this section, we describe the numerical approach for simulating the flows over a sphere and a spheroid. The governing equations (\ref{eq:NS}) are discretized and solved using a fractional-step approach with a local volume-flux formulation on a staggered curvilinear grid \citep{wang2019discrete,You2019tbl}. The advection terms are discretized using the Adams-Bashforth scheme, and the Crank-Nicolson scheme is adopted for the diffusion terms. The pressure Poisson equation is solved using bi-conjugate gradient stabilized method (BICGSTAB) with an algebraic multigrid preconditioner provided by \emph{Hypre} \citep{falgout2002hypre}. These algorithms are first implemented and validated on a single-block domain. For the simulation of flow over a bluff body, the fluid domain is decomposed into six blocks, as shown in figure \ref{fig: domain grid}. Within each block the advection and diffusion terms are discretized using the single-block description noted above, and on the block boundaries the diffusion terms are discretized using the Adams-Bashforth scheme. The pressure fields in all blocks are solved globally due to the ellipticity of the pressure Poisson equation. 

The computational domains and grids for the flows over the sphere and spheroid are shown in figure \ref{fig: domain grid}. The fluid domain in the former case is formed by two concentric sphere surfaces with $R_1 = 0.5$ and $R_2 = 15$, and divided into six blocks for the structured multi-block flow solver. No-slip and free-stream boundary conditions are imposed for the velocity field at the inner and outer sphere surfaces,  respectively. Each block is discretized into a structured curvilinear mesh. The number of grid points on the interface between two blocks is the same on the two sides, which constrains the number of grid points within different blocks.  On accounting of these constraints, there are four independent number of grid points $N_x, N_y, N_z, N_w$ that can be specified on this multi-block grid, which are shown in figure \ref{fig:flow over bluff body}. The flow domain, mesh, and boundary conditions for the flow over the prolate spheroid are similarly configured. The fluid domain is formed between two concentric spheroids with their axes aligned, decomposed into six blocks, and discretized on a structured curvilinear mesh similar to the sphere cases. The aspect ratio of the inner spheroid is $a/b=6\colon1$, as shown in figure \ref{fig:flow over bluff body} (b). The outer spheroid has aspect ratio close to unity and radii equal to $16.44$ and $16.17$, with its major axes aligned with the inner one.  The geometry, mesh parameters, and Reynolds numbers are reported in table \ref{table:numerical setup}. The case designations `SL', `ST', `PT' refer to laminar flow over the sphere, turbulent flow over the sphere, and turbulent flow over the prolate spheroid. 

Visualizations of the vortical structures in the three flows are shown in figure \ref{fig:vortical structures}. The distinct characteristics of the vortical structures emphasize the different purposes of these three examples. The simplicity of case `SL', reflected by figure \ref{fig:vortical structures}(a), enables a concise demonstration of the theoretical elements discussed in \S\S\ref{subsec:Flow over bluff body}-\ref{subsec:Detailed Josephson-Anderson relation}. Vortex shedding and a turbulent wake are introduced by considering case `ST' (figure \ref{fig:vortical structures}(b)), where a statistical description of the vorticity dynamics is required. The pair of vortices above the spheroid in case `PT' (figure \ref{fig:vortical structures}(c)) are closely related with the three-dimensional separation on the wall. The implication for vorticity transport by these vortices and the underlying separation patterns are the focus of case `PT'. 

The J-A relation (\ref{eq:JA}) and the Huggins flux tensor (\ref{eq:Huggins tensor}) are numerically evaluated for the three simulated flows described earlier. The vorticity is computed using a finite-volume scheme on generalized curvilinear coordinate system \citep{Rosenfeld1991}.  This vorticity is then used to evaluate the advection term in the J-A relation and the advective vorticity flux. The vorticity diffusion vector is computed based on the identity 
$\nabla^2\boldsymbol{u}=-\nabla\times\boldsymbol{\omega}$, where the Laplacian of the velocity field is obtained from the right-hand side of the momentum equation in the Navier-Stokes solver. This procedure ensures that the numerical evaluation of vorticity diffusion is consistent with the momentum balance enforced in the solver.

\begin{figure}
    \centering
    \includegraphics[width=0.7\textwidth]{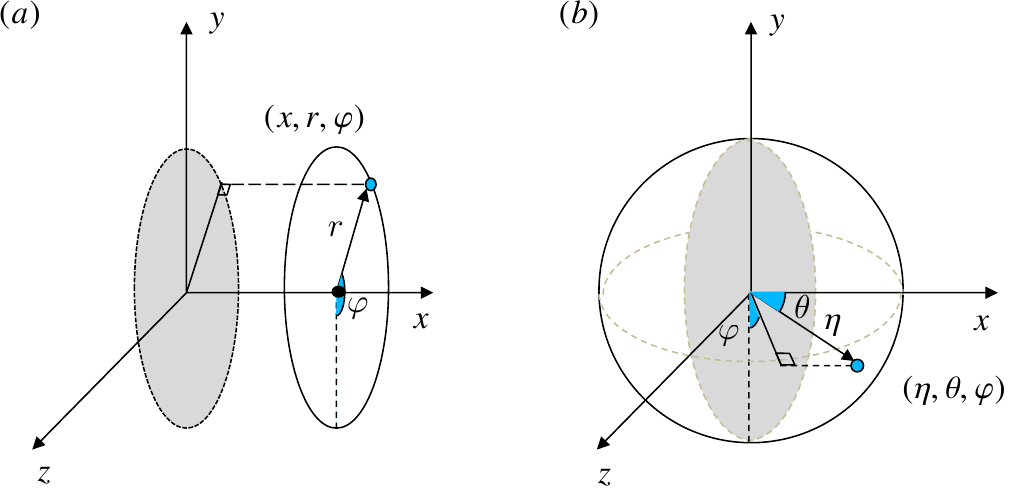}
    \caption{Schematics of the $(a)$ cylindrical and $(b)$ spheroid coordinate systems which are adopted in the analysis of vorticity fluxes. $(a)$ The $x$-axis is aligned with the polar coordinate, the azimuthal angle is denoted by $\varphi$, and the radial coordinate is $r$. $(b)$ The polar angle $\theta$ is formed by the polar axis ($x$-axis) and the radial vector. The length of the radial vector is denoted by $\eta$. The azimuthal angle $\varphi$ is formed with respect to the $y$-direction. }
    \label{fig:coordinates}
\end{figure}

Both cylindrical and spherical coordinates will be utilized to present the numerical results, and are schematically shown in figures \ref{fig:coordinates}.  The origins of both coordinates are placed at the center of the sphere for cases `SL' and `ST', and the free-stream velocity lies along the $x$ axis. 
For the flow over the spheroid, the center of the spheroid is located at $(x,y,z)=(\frac{a}{2},0,0)$, and the major axes lies with $x$-axis. The free-stream velocity is $\boldsymbol{U}=(\cos(\alpha),\sin(\alpha),0)$, where $\alpha$ is the incidence angle.

\section{Vorticity dynamics in flow over a sphere}\label{sec:sphere results}

In this section, the vorticity dynamics and its connection to the rate of work exerted on the fluid by the drag force for flow over a sphere is quantitatively studied using the Josephson-Anderson relation. Although the geometry is simple, the flow exhibits rich physical phenomena at different Reynolds numbers. We choose $\Rey=200$ (case `SL') as the first example because the flow at this Reynolds number is sufficient to build a physical picture of vorticity transport. We then proceed to study the vorticity transport in the unsteady, impulsively started, turbulent wake (case `ST') $\Rey=3700$.

\subsection{Laminar flow over a sphere}\label{subsection:Re200}

The flow over a sphere at Reynolds number $\Rey=200$ is steady and axis-symmetric. The forebody of the sphere is wrapped with an attached laminar boundary layer that separates at approximately $\theta = 62^{\circ}$ due to the adverse pressure gradient and curvature. A steady cylindrical shear layer that wraps the recirculation region forms behind the sphere. Farther downstream, the flow recovers from the defect profile towards the uniform free-stream velocity.

\begin{figure}
    \centering
    \includegraphics[width=0.4\textwidth]{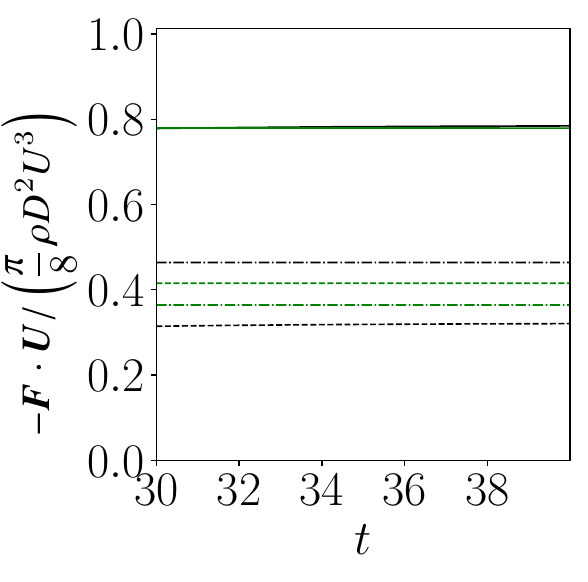}
    \caption{Time history of drag coefficient for case SL from integration of wall pressure and shear stress Josephson-Anderson relation. \fullgreen{}: total drag work evaluated by surface integration of pressure and friction, \dashedgreen{}: pressure work, \chaingreen{}: friction work, \fullblack{}: total drag work evaluated from the J-A relation, \dashedblack{}: total advective contribution $\int_{\Omega}\boldsymbol{u}_{\phi}\cdot\left(-\boldsymbol{u}\times\boldsymbol{\omega}\right)\mathrm{d}V$, \chainblack{}: total viscous contribution $\int_{\Omega}\boldsymbol{u}_{\phi}\cdot\left(\nu\boldsymbol{\nabla}\times\boldsymbol{\omega}\right)\mathrm{d}V$. }
    \label{fig:sphere200_history}
\end{figure}

\begin{figure}
    \centering
    \includegraphics[width=0.8\textwidth]{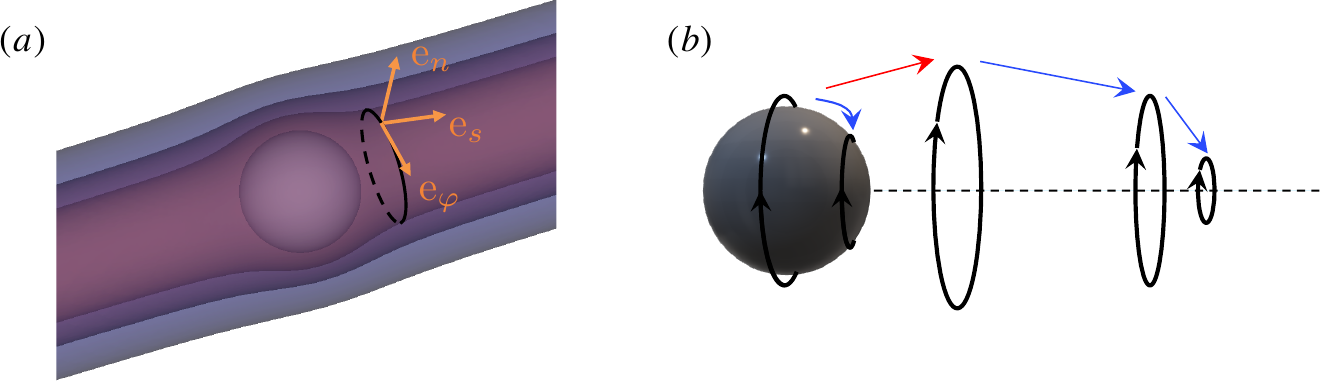}
    \caption{Schematics of potential flow and vorticity transport direction. (a) The translucent surfaces represent the iso-surfaces of $\psi=\{-0.1,-0.2,-0.4\}$. The vectors $\boldsymbol{e}_{\phi}$,$\boldsymbol{e}_{s}$,$\boldsymbol{e}_{n}$ is a set of local orthogonal coordinates. (b) The solid black lines with arrows represent the azimuthal vortex rings. The arrows with red and blue colors represent the outward and inward transport of vorticity crossing the iso-surface of $\psi$.   }
    \label{fig: sphere potential schematics}
\end{figure}

\begin{figure}
    \centering
    \includegraphics[width=1.0\textwidth]{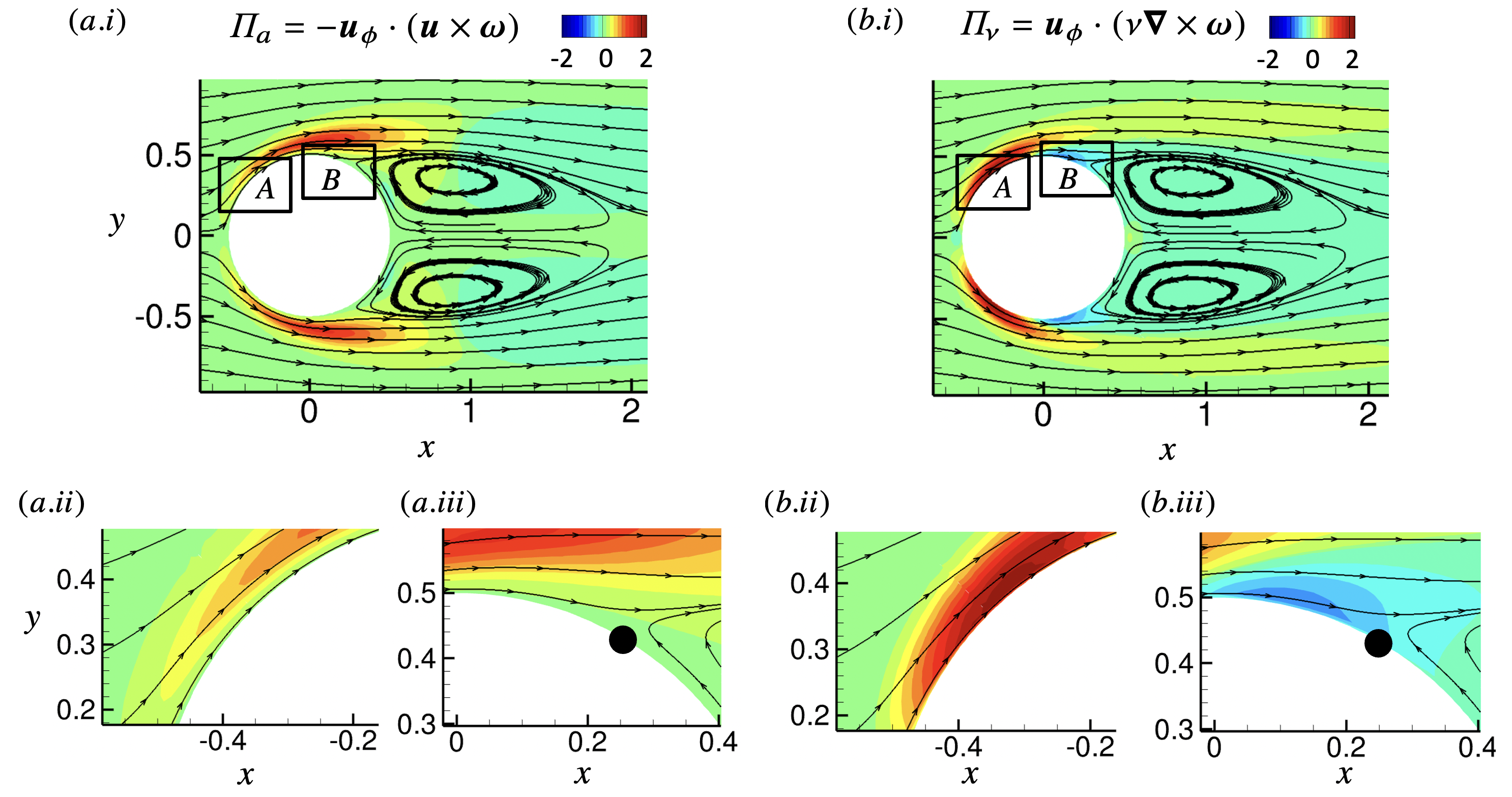}
    \caption{Two dimensional contour of instantaneous value of vorticity fluxes for case SL overlapped with streamlines. Coloured contours show $(a)$ $\Pi_{a}$ and $(b)$ $\Pi_{\nu}$. Panels $(a.ii)$ and $(b.ii)$ are zoomed-in view of boxed region $A$, and panels $(a.iii)$ and $(b.iii)$ are zoomed-in view of boxed region $B$.} 
    \label{fig: sphere200 instantaneous J-A contour}
\end{figure}

We compare two approaches to evaluating the power injection into the fluid, $-\boldsymbol{F} \cdot \boldsymbol{U}$. First, the drag force is evaluated by integrating the pressure and shear stress over the sphere surface using equation (\ref{eq:drag force}). In the second approach, we evaluate the J-A relation (\ref{eq:JA}). In the latter case, the potential flow is given by the analytical expression,
\begin{equation}
    \phi=U\left(\eta+\frac{R_1^{3}}{2 \eta^{2}}\right) \cos \theta, \qquad\psi=\frac{1}{2} U \eta^{2} \sin ^{2} \theta\left(1-\frac{R_1^{3}}{\eta^{3}}\right),  
\end{equation}
where $\eta$ is the radial distance in the spherical coordinate system. The comparison in figure \ref{fig:sphere200_history} demonstrates that the two approaches to evaluating $-\boldsymbol{F} \cdot \boldsymbol{U}$ agree, both yielding equal time histories.  
The ingredients in terms of the wall-pressure and wall-shear stress, as well as from the vorticity fluxes by advection and diffusion, are also visualized in the same figure. The magnitude of pressure and friction drag are similar and remain constant through time in this case. Instead of dividing the drag force into form and friction drag, the J-A relation expresses the rate of drag work into advective $\int_{\Omega}\Pi_{a}\mathrm{d}V$ and diffusive $\int_{\Omega}\Pi_{\nu}\mathrm{d}V$ terms. The advective part is most closely related to form drag since the advection of vorticity is accompanied with a pressure gradient in the transverse direction. The diffusive part, on the other hand, is most closely related to the friction drag.  These four components of drag have similar magnitudes in figure \ref{fig:sphere200_history} due to the low Reynolds number considered here. Therefore, it can be anticipated that the comparison between the two approaches to compute drag, and the differences between two contributions within each approach, will be more evident in the turbulent case `ST'.

For axisymmetric flows, the J-A relation (\ref{eq:JA1}) can be further re-written using the flux of the azimuthal vorticity.  We denote the unit-vector in the azimuthal direction as  $\boldsymbol{e}_{\varphi}$, and that along the potential-flow streamlines as $\boldsymbol{e}_{s}$.  Therefore, the unit-vector $\boldsymbol{e}_{n}$ normal to an iso-surface of the potential-flow streamfunction is given by $\boldsymbol{e}_{n} = \boldsymbol{e}_{\varphi} \times \boldsymbol{e}_{s}$, as visualized in figure \ref{fig: sphere potential schematics}$(a)$. The J-A relation is then re-written as: 
\allowdisplaybreaks
\begin{eqnarray}
    \label{eq:JA_axisymmetric}
    -\boldsymbol{F} \cdot \boldsymbol{U} &=& - \int\mathrm{d}J\int\left(\boldsymbol{u}\times\boldsymbol{\omega}-\nu\boldsymbol{\nabla}\times\boldsymbol{\omega}\right)\mathrm{d}\boldsymbol{l}\nonumber \\ 
    &=& -\int\mathrm{d}J\int\left(\boldsymbol{u}\times\boldsymbol{\omega}-\nu\boldsymbol{\nabla}\times\boldsymbol{\omega}\right)\cdot\boldsymbol{e}_{s} \mathrm{d}l\nonumber\\ 
    &=& - \int\mathrm{d}J\int\left(\boldsymbol{u}\times\boldsymbol{\omega}-\nu\boldsymbol{\nabla}\times\boldsymbol{\omega}\right)\cdot\left(\boldsymbol{e}_{n}\times\boldsymbol{e}_{\varphi}\right) \mathrm{d}l\nonumber\\ 
    &=& - \int\mathrm{d}J\int\boldsymbol{e}_{\varphi}\cdot\left(\left(\boldsymbol{u}\times\boldsymbol{\omega}-\nu\boldsymbol{\nabla}\times\boldsymbol{\omega}\right)\times\boldsymbol{e}_{n}\right)  \mathrm{d}l \nonumber\\ 
    &=& - \int\mathrm{d}J\int\boldsymbol{e}_{\varphi}\cdot\boldsymbol{\Sigma}\cdot\boldsymbol{e}_{n} \mathrm{d}l \nonumber\\
    &=& - \int\mathrm{d}J\int\boldsymbol{e}_{n}\cdot\boldsymbol{\Sigma}\cdot\left(-\boldsymbol{e}_{\varphi}\right) \mathrm{d}l, 
    \label{eq:JAaxsisym}
\end{eqnarray}
where $\mathrm{d}\boldsymbol{l}=\boldsymbol{e}_{s} \mathrm{d}l$ is the vector line element pointing along potential velocity, and $\mathrm{d}l$ is the length of this line element. Figure \ref{fig: sphere potential schematics}$(b)$ is a schematic of vortex loops being generated from the sphere surface and transported into the wake. The red and blue arrows represent vortex loops crossing the potential streamline outward and inward, and correspond to the positive and negative rates of work by drag, respectively. The expression $\boldsymbol{e}_{n}\cdot\boldsymbol{\Sigma}\cdot\left(-\boldsymbol{e}_{\varphi}\right)$ represents the flux of negative azimuthal vorticity in the $\boldsymbol{e}_{n}$ direction. In this form of the J-A relation, the rate of work done by the drag force is represented as a weighted integral of azimuthal vorticity flux crossing the iso-surface of the stream function. The vorticity flux tensor $\boldsymbol{\Sigma}=\boldsymbol{\Sigma}_a+\boldsymbol{\Sigma}_{\nu}$, can be simplified into the following forms: 
\begin{equation}
    \boldsymbol{\Sigma}_a = \begin{pmatrix}
0 & 0 & u_x\omega_{\varphi}\\
0 & 0 & u_r\omega_{\varphi}\\
-u_x\omega_{\varphi} & -u_r\omega_{\varphi} & 0\\
\end{pmatrix}, \qquad
\boldsymbol{\Sigma}_{\nu} = \begin{pmatrix}
0 & 0 & -\nu\frac{\partial\omega_{\varphi}}{\partial x}\\
0 & 0 & -\nu\frac{\partial\omega_{\varphi}}{\partial r}\\
\nu\frac{\partial\omega_{\varphi}}{\partial x} &\nu\frac{\partial\omega_{\varphi}}{\partial r}  & 0
\end{pmatrix}. 
\end{equation}
The flux of azimuthal vorticity can be obtained by dotting the above expressions with the azimuthal unit-vector $\boldsymbol{e}_{\varphi}$: 
\begin{equation}
        \boldsymbol{\Sigma}_{a}\cdot \boldsymbol{e}_{\varphi} = \begin{pmatrix}
u_x \omega_{\varphi}\\
u_r \omega_{\varphi} \\
0\\
\end{pmatrix},\qquad
    \boldsymbol{\Sigma}_{\nu}\cdot \boldsymbol{e}_{\varphi} = \begin{pmatrix}
-\nu\frac{\partial\omega_{\varphi}}{\partial x} \\
-\nu\frac{\partial\omega_{\varphi}}{\partial r} \\
0\\
\end{pmatrix}. 
\end{equation}
The explicit forms of $\boldsymbol{\Sigma}_{a}\cdot \boldsymbol{e}_{\varphi}$ and $\boldsymbol{\Sigma}_{\nu}\cdot \boldsymbol{e}_{\varphi}$ obtained above can be clearly interpreted as the advection of $\omega_{\varphi}$ by the velocity field and the down-gradient diffusion of $\omega_{\varphi}$ by viscosity, respectively. The portion of these fluxes which contributes to the action of drag is the projection onto $\boldsymbol{e}_{n}$, i.e.\,the flux normal to the potential flow, as shown in the last expression of equation (\ref{eq:JA_axisymmetric}).

The advective contribution 
$\Pi_a(\boldsymbol{x})= -\boldsymbol{u}_{\phi}\cdot\left(\boldsymbol{u}\times\boldsymbol{\omega}\right) =|\boldsymbol{u}_{\phi}|\boldsymbol{e}_{n} \cdot\boldsymbol{\Sigma}_a\cdot\left(-\boldsymbol{e}_{\varphi}\right)$
and viscous contribution 
$
\Pi_{\nu}(\boldsymbol{x}) =|\boldsymbol{u}_{\phi}|\boldsymbol{e}_{n} \cdot\boldsymbol{\Sigma}_{\nu}\cdot\left(-\boldsymbol{e}_{\varphi}\right) = \boldsymbol{u}_{\phi}\cdot\left(\nu\boldsymbol{\nabla}\times\boldsymbol{\omega}\right)
$ 
are visualized in figure \ref{fig: sphere200 instantaneous J-A contour}, overlapped with flow streamlines. 
In the near-wall region $A$, the viscous contribution dominates inside the boundary layer, while the advective contribution is close to zero near the wall due to the no-slip condition. The positive viscous flux contributes to drag power injection in a region along the wall in the attached boundary layer, which is associated with the outward  transport of the negative azimuthal vorticity that is introduced into the fluid domain by the favorable pressure gradient. 
The vortex force in the near-wall region $A$ drives the flow from potential to vortical, and results in a transfer from the interaction energy to vortical energy (interpretation from equation (\ref{eq:interaction energy})). A weak negative viscous contribution appears near $\theta=90^{\circ}$ at the wall prior to separation, and extends downstream as shown in region $B$ in figure \ref{fig: sphere200 instantaneous J-A contour}. This negative flux, which reduces the rate of drag work, is due to the change in pressure gradient from favorable to adverse and the associated wall vorticity flux $\sigma$. After the negative azimuthal vorticity diffuses into the domain, it is advected along the detached boundary layer crossing the potential streamlines outwards.  This flux drives the flow away from the ideal potential flow, and thus produces a positive contribution to drag power. In the laminar wake, the streamwise velocity recovers from a deficit profile to the free-stream potential flow. The negative contribution to drag power near the wake centerline indicates that the vortex force drives the flow from vortical to potential, and that the kinetic energy is transferred from vortical to interaction energy.

\begin{figure}
    \centering
    \includegraphics[width=1.0\textwidth]{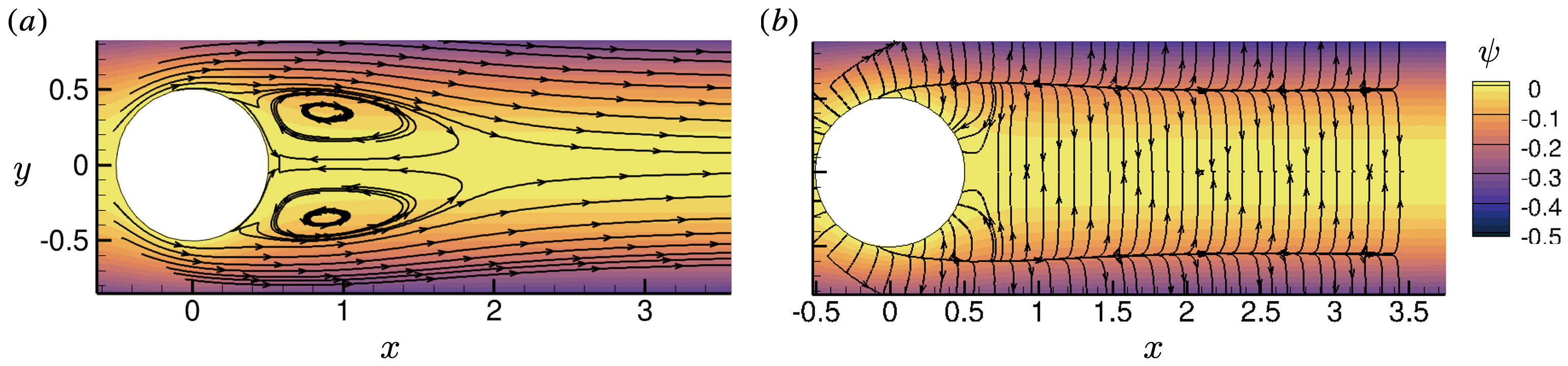}
    \caption{Visualization of vorticity flux vector field. The lines in the panels are tangent to the following vector fields (a) $\boldsymbol{\Sigma}_a\cdot\boldsymbol{e}_{\varphi}$ and (b)  $\boldsymbol{\Sigma}_{\nu}\cdot\boldsymbol{e}_{\varphi}$. The color contours represent the stream-function $\psi$ of the background potential flow. }
    \label{fig:sphere200 flux trace}
\end{figure}

The link between the vorticity flux and the rate of drag work underscores the importance of a quantitative description of vorticity transport, which is given by the Huggins flux tensor. The evolution equation for a general vorticity component $\omega_{\xi} = \boldsymbol{\omega}\cdot \boldsymbol{e}_{\xi}$ can be derived by dotting a constant unit-vector $\boldsymbol{e}_{\xi}$ with the vorticity equation (\ref{eq:Helmholtz}), 
\begin{equation}\label{eq:planar conservation law}
    \frac{\partial \omega_{\xi}}{\partial t} + \boldsymbol{\nabla} \cdot \left(\boldsymbol{\Sigma}\cdot\boldsymbol{e}_{\xi}\right) = 0. 
\end{equation}
The above relation is a planar conservation law for the transport of $\omega_{\xi}$ within a plane $S_{\boldsymbol{e}_{\xi}}$ that is normal to the unit-vector $\boldsymbol{e}_{\xi}$. Due to the antisymmetry of the flux tensor $\boldsymbol{\Sigma}$, the vorticity flux of $\omega_{\xi}$ in the direction $\boldsymbol{e}_{\xi}$ vanishes, since $\boldsymbol{e}_{\xi} \cdot \boldsymbol{\Sigma}\cdot \boldsymbol{e}_{\xi} = 0$. In other words, in absence of boundaries, the vorticity component $\omega_{\xi}$ is conserved within the plane $S_{\boldsymbol{e}_{\xi}}$, as can be seen by integrating (\ref{eq:planar conservation law}) on the plane $S_{\boldsymbol{e}_{\xi}}$. The vector field $\boldsymbol{\Sigma}\cdot\boldsymbol{e}_{\xi}$ is the flux of $\omega_{\xi}$, and serves as an informative visualization of its transport. 

In the present case of laminar separation (SL), the vorticity field is comprised of an azimuthal component only. We therefore consider the azimuthal unit-vector $\boldsymbol{e}_{\varphi}$ and visualize the vector field $\boldsymbol{\Sigma}\cdot\boldsymbol{e}_{\varphi}$. The streamtraces parallel to the vector fields $\boldsymbol{\Sigma}_a\cdot\boldsymbol{e}_{\varphi}$ and $\boldsymbol{\Sigma}_{\nu}\cdot\boldsymbol{e}_{\varphi}$ are visualized in figure \ref{fig:sphere200 flux trace}. 
The advective flux $\boldsymbol{\Sigma}_a\cdot\boldsymbol{e}_{\varphi}$ does not start or end at the wall, which implies that this flux only redistributes vorticity within the volume, and thus changes the vorticity magnitude locally without altering the volume integral of vorticity. 
In contrast, the viscous flux is mainly responsible for the outward diffusion of vorticity from the wall into the fluid and its annihilation at the wake centerline. This interpretation is consistent with the classical Eulerian picture of vorticity transport: 
Negative azimuthal vorticity is generated at the wall by pressure gradient, and vorticity diffuses into boundary layer.  Within the cylindrical shear layer, vorticity is mostly advected.
The azimuthal vorticities from different azimuthal angles cancel each other at the wake centerline. The inward transport of vorticity observed in figure \ref{fig:sphere200 flux trace} corresponds to the anti-drag region appearing in figure \ref{fig: sphere200 instantaneous J-A contour} near the wake centerline. The rate of vorticity annihilation, defined by the vorticity flux $\left(\boldsymbol{\Sigma}_\nu\right)_{r\varphi }$, can be related to streamwise total pressure gradient by simplifying equation (\ref{eq:mom_S}) at the wake centerline \citep{terrington2021generation}: 
\begin{equation}\label{eq: momt_wake_laminar}
    \left(\boldsymbol{\Sigma}_\nu\right)_{r\varphi}=-\nu\frac{\partial\omega_{\varphi}}{\partial r}=\frac{\partial h}{\partial x}. 
\end{equation}
The recovery of total pressure is accompanied by inward transport of azimuthal vorticity towards the wake centerline. This inward flux only contains a viscous contribution at the centerline, meaning that viscous effect is solely responsible for the annihilation of azimuthal vorticity.

\subsection{Impulsively started flow over a sphere: multiple two dimensional separations}
\label{subsection:Re3700 starting}

\begin{figure}
    \centering
    \includegraphics[width=1.0\textwidth]{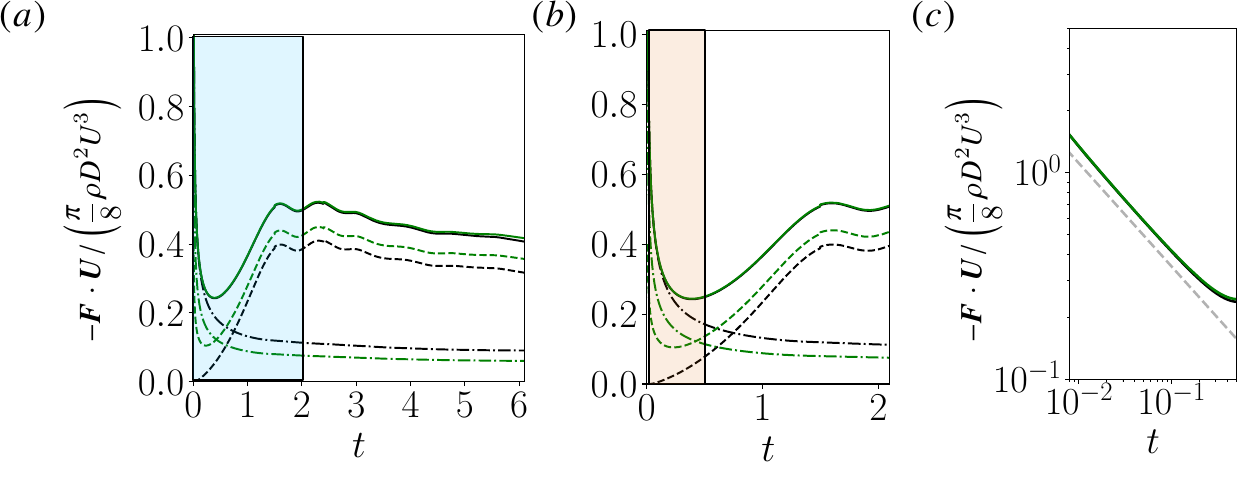}
    \caption{($a$) Time history of drag coefficient for case ST from integration of wall-pressure and wall-shear stress, and from the Josephson-Anderson relation.  Panel ($b$) is a zoomed-in view of $(a)$ during $0\leq t \leq 2$.  
    \fullgreen{}: Total drag work evaluated by surface integration of pressure and friction, \dashedgreen{}: pressure work, \chaingreen{}: friction work, \fullblack{}: total drag work evaluated from the J-A relation, \dashedblack{}: total advective contribution $\int_{\Omega}\Pi_a\mathrm{d}V$, \chainblack{}: total viscous contribution $\int_{\Omega}\Pi_\nu\mathrm{d}V$. Panel $(c)$ is a further zoomed-in view of $(b)$ in $0.08\leq t\leq0.5$ in log scale, and including (\dashedgray{}) the Basset-Boussinesq force $F_B$. }
    \label{fig:sphere3700 starting history}
\end{figure}
\begin{figure}
    \centering
    \includegraphics[width=1.0\textwidth]{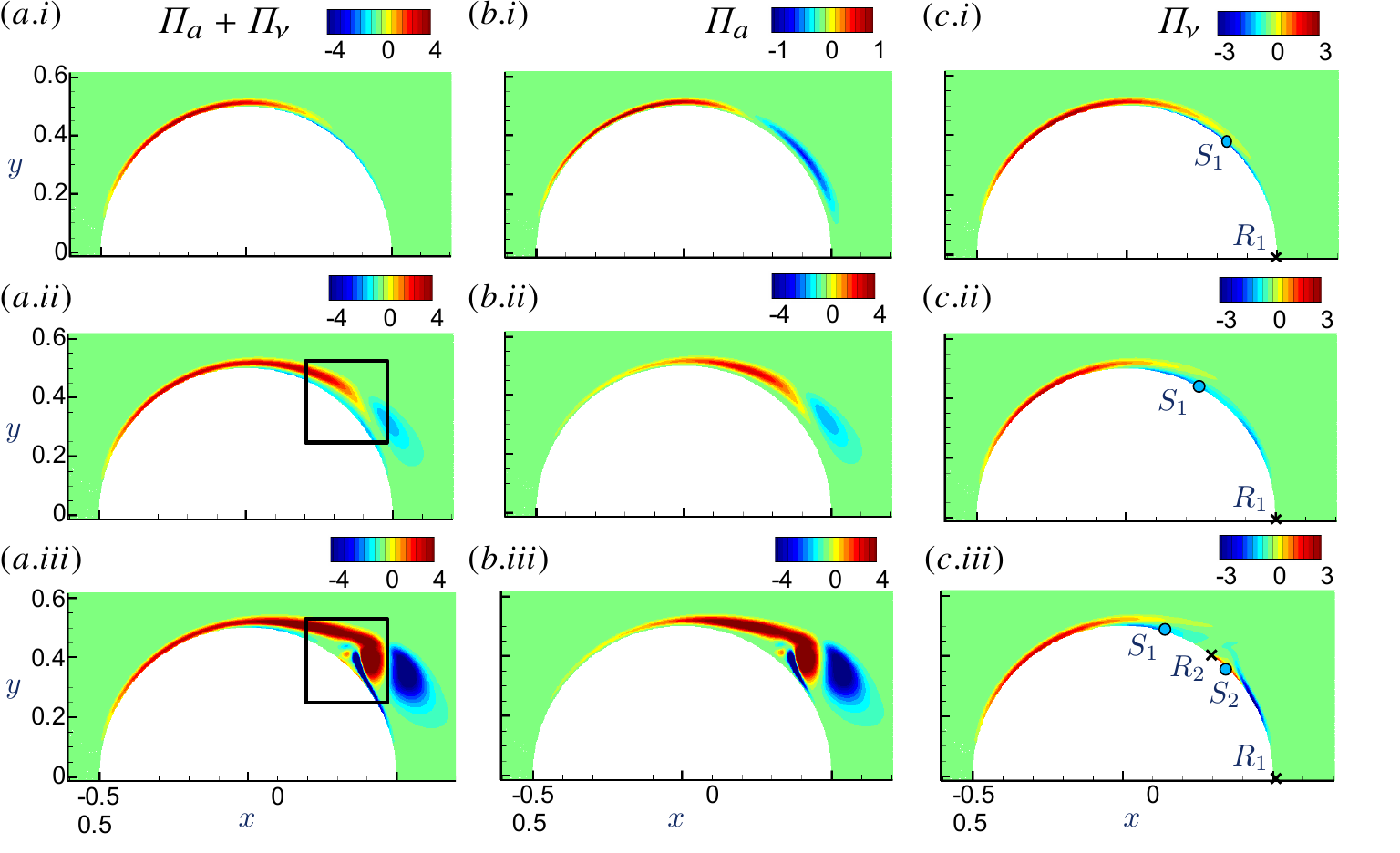}
    \caption{Contours of instantaneous vorticity fluxes for case ST at (i-iii) $t=\{0.3, 0.9, 1.5\}$.  The region enclosed in the black box is visualized in figure \ref{fig: sphere3700 starting J-A pres streamline}}
    \label{fig: sphere3700 starting instantaneous J-A contour}
\end{figure}
\begin{figure}
    \centering
    \includegraphics[width=0.7\textwidth]{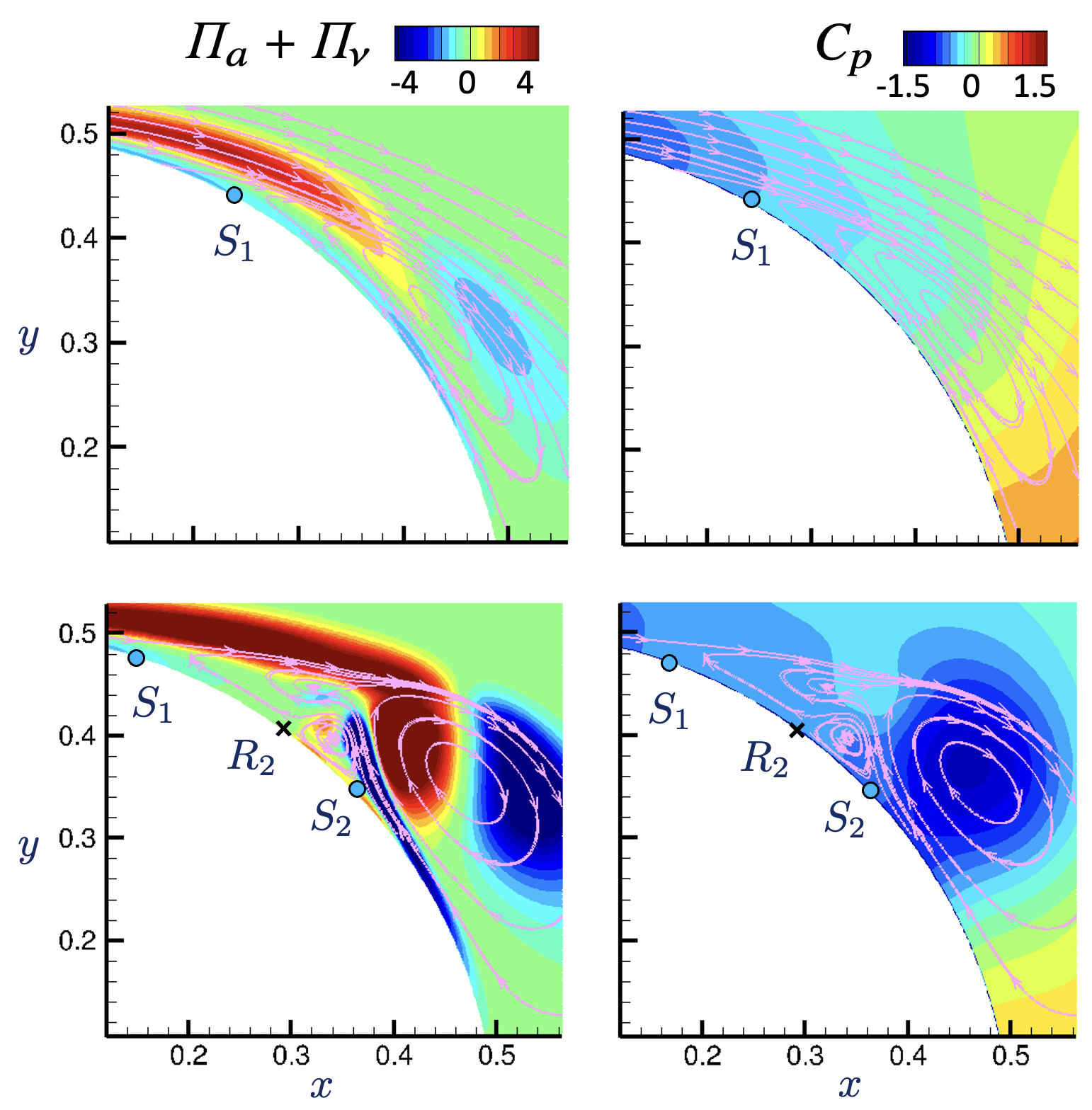}
    \caption{Two dimensional contour of instantaneous value of drag contribution and pressure coefficients for case ST. The top and bottom row correspond to  $t=\{0.9,1.5\}$. }
    \label{fig: sphere3700 starting J-A pres streamline}
\end{figure}

The flow over a sphere at $\Re=3700$ is simulated and is designated `ST' (see table \ref{table:numerical setup}). The simulation starts from a initial condition with uniform free-stream velocity in the fluid domain everywhere. The time history of the terms in the J-A relation, as well as the rates of form and friction drag work are visualized in figure \ref{fig:sphere3700 starting history}. The total drag force drops from an unbounded value at $t=0$ to a finite value at later times due to the $t^{-1/2}$ singularity of the drag force at the initial time \citep{mei1996flow}, from laminar boundary-layer scaling. 
This $t^{-1/2}$ scaling can be related to the Basset-Boussinesq force \citep{basset1888treatise, boussinesq1903theorie}, which accounts for the history effect in unsteady Stokes flow around a sphere. The relative acceleration between the body and the freestream influences the development of the boundary layer, which subsequently affects the drag force. This history effect is captured in the Basset-Boussinesq force, 
\begin{equation}
    F_B(t) = 6 a^2 \sqrt{\pi \nu} \int_{-\infty}^t  \frac{\ddot{X}\left(t^{\prime}\right)}{\sqrt{t-t^{\prime}}}\mathrm{~d} t^{\prime}, 
\end{equation}
where $\ddot{X}\left(t^{\prime}\right)$ is the acceleration of the body at $t'$. In the present case of an impulsively started flow over a sphere, the acceleration is a delta function at $t=0$. Therefore, for $t>0$ the Basset-Boussinesq force is:
\begin{equation}
    F_B(t) =6 a^2U\sqrt{\frac{ \pi \nu}{t}}.
\end{equation}
This force decays proportionally to $t^{-1/2}$, which directly corresponds to the initial decay observed in the total drag. 
Figure \ref{fig:sphere3700 starting history}(c) shows the initial decay of the drag force for the impulsively started flow over a sphere and the corresponding Basset-Boussinesq force $F_B$. The $t^{-{1/2}}$ decay of the drag force agrees well with the decay of $F_B$. 
Note that although at $t\to0^{+}$, the thickness of the boundary layer over the sphere approaches zero and most of the flow field is identical to potential flow, the total pressure drag does not approach zero. The divergence of the Lamb vector $\nabla\cdot(\boldsymbol{u}\times\boldsymbol{\omega})$ is the source term in the Poisson equation of the enthalpy function $h=\frac{p}{\rho}+\frac{1}{2}\vert\boldsymbol{u}\vert^2$. The thin singular vortex sheet on the sphere wall exerts a non-local effect on the pressure field which induces a non-zero form drag even at the initial time. The asymmetry of the pressure profile with respect to the polar angle at the initial time was verified by \citet{dennis1972numerical} and by \citet{wang1969impulsive}, and was similarly attributed by \citet{kang1988drag} to the viscous generation of a vortex sheet on the surface for flow over a spherical bubble. The total drag from the J-A relation agrees well with the combination of form and friction drags. A distinction arises between the conventional division of drag into friction/form parts and the J-A division into advection/diffusion parts at $t\to 0^+$. The pressure and friction drags are both non-zero at the initial time, meanwhile J-A relation attributes all of the initial drag power injection to the viscous flux through the wall. 
The advection contribution is zero at $t \to 0$, since the Lamb vector $\boldsymbol{u}\times\boldsymbol{\omega}$ is orthogonal to the potential-flow direction in the vortex sheet on the wall. At this initial time, the J-A interpretation of drag force in terms of advective and diffusive fluxes of vorticity is more instructive compared to the conventional decomposition into shear stress and pressure. Specifically, diffusion of vorticity is the sole relevant actor at the initial time, while the shear stress and pressure are byproducts of the action of the instantaneous diffusive flux. The viscous contribution drops rapidly during the initial development stage since the viscous flux reduces as the boundary layer grows. The advective contribution increases gradually from zero because of the detachment of the boundary layer and vortex formation on the back of the sphere. At later times, $t>1$, the viscous contribution in the J-A relation reduces, while the advection part increases and dominates the total rate of work by drag due to the detachment of vorticity from the near-wall region.

The spatial contributions to drag power injection $\Pi_a$ and $\Pi_\nu$ are visualized in figure \ref{fig: sphere3700 starting instantaneous J-A contour} at three different times, $t=\{0.3,0.9,1.5\}$. At the first instant, all of the vorticity in this flow is confined to the thin boundary layer on the sphere surface. Most of the drag originates from the viscous effect within the surface layer, where the vorticity transport pushes the flow away from ideal to vortical and produces drag force. A thin reverse-flow layer is already forming at $\varphi>90^{\circ}$ due to the adverse pressure gradient. The locations $S_1$ and $R_1$ represent the primary separation and re-attachment, which are lines of zero friction along the azimuthal direction. At $t=0.9$, the total advection and viscous contributions become comparable. The advection part $\Pi_a$ becomes stronger due to the detachment of the primary boundary layer from the geometry and the formation of an axisymmetric primary vortex, as also shown in figure \ref{fig: sphere3700 starting J-A pres streamline}. The separation location $S_1$ moves upstream due to the adverse tangential pressure gradient. At $t=1.5$, the primary vortex starts to detach from the solid wall, with a low pressure region forming at the vortex core. The secondary boundary layer beneath the vortex starts to separate due to a pressure gradient that is adverse for the secondary layer but favorable for the primary flow. A secondary vortex forms beneath the primary one between $S_2$ and $R_2$ which appear in a pair due to the topological constraints on singular points on the surface \citep{tobak1982topology}.

\subsection{Turbulent flow over a sphere}\label{subsection:Re3700}

\begin{figure}
    \centering
    \includegraphics[width=0.9\textwidth]{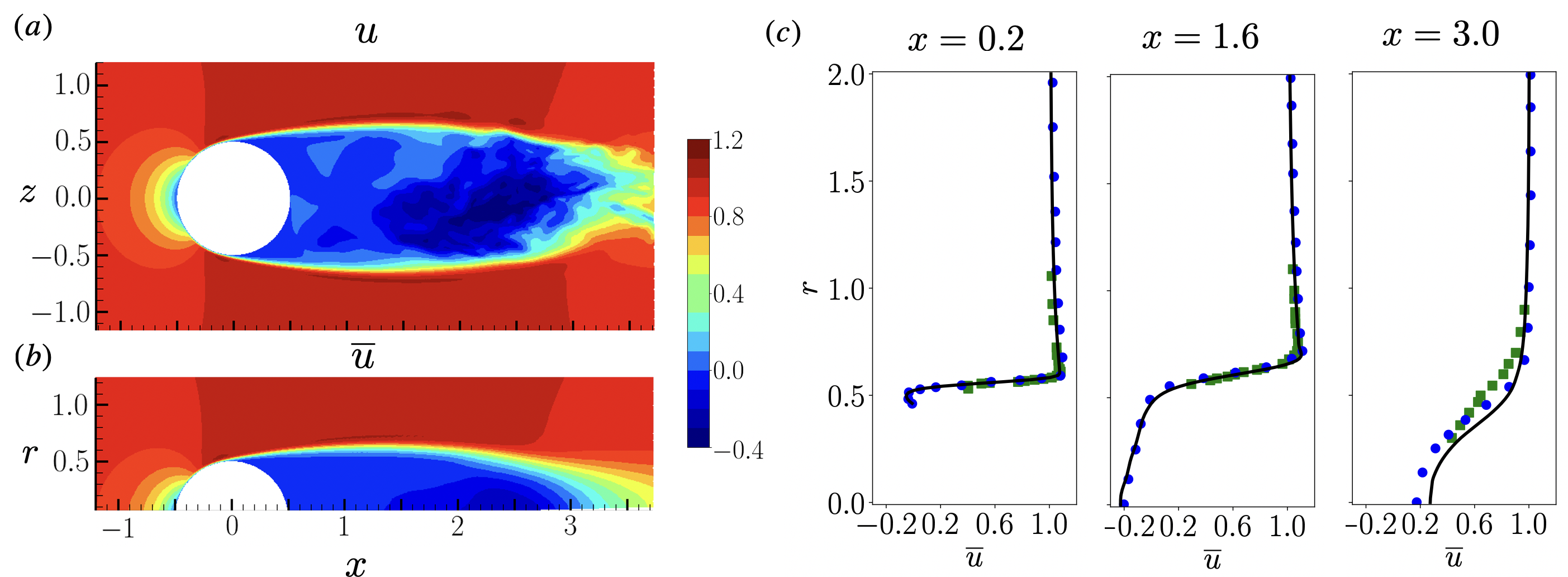}
    \caption{The streamwise velocity for case ST. (a) A snapshot of instantaneous $u$-velocity around the sphere during the statistically stationary stage. (b) Mean streamwise velocity, $\overline{u}$. (c) Comparison of the mean-velocity profiles from (\fullblack{}) the present simulations with (\mycircle{blue} \mycircle{blue} \mycircle{blue}) previous DNS by \citet{rodriguez2011direct} and (\mysquare{black!40!green} \mysquare{black!40!green} \mysquare{black!40!green}) experimental data by \citet{kim1988observations}. The $\overline{u}$ profiles are plotted along the radial direction in a cylindrical coordinate at streamwise locations $x=\{0.2,1.6,3.0\}$}
    \label{fig:sphere3700_U}
\end{figure}

After the starting stage discussed in \S\ref{subsection:Re3700 starting}, the wake continues to extend downstream. The detached cylindrical shear layer breaks down to turbulence, and a statistically stationary state is established within the simulation domain after a sufficiently long time. The transient time from the initial condition to the stationary stage was approximately $150$ advective time units. Flow statistics were evaluated by averaging over $100$ time units and in the azimuthal direction. At this Reynolds number, although the boundary layer on the sphere is still laminar, the detached shear layer develops instability at approximately $x=2$ \citep{rodriguez2011direct,yun2006vortical} where localized azimuthal vortex roll-up starts to appear. At the end of the separation bubble, corrugated structures along the azimuthal direction develop and ultimately the wake breaks down to turbulence, which is observed in the instantaneous visualization of streamwise velocity in figure \ref{fig:sphere3700_U}$(a)$. The mean streamwise velocity field, as well as its comparison with previous DNS and experimental data, are show in figure \ref{fig:sphere3700_U}$(b,c)$. We focus on the physical interpretation of the Josephson-Anderson relation and the vorticity dynamics in the presence of the turbulent wake.
\begin{figure}
    \centering
    \includegraphics[width=0.7\textwidth]{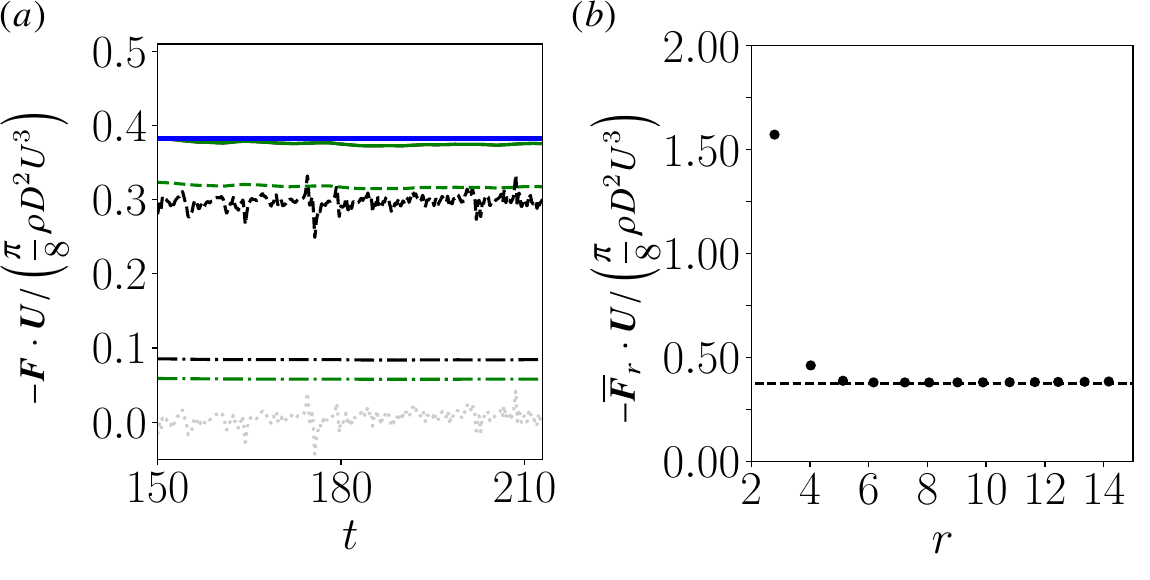}
    \caption{(a): Time history of the drag coefficient for case `ST' from integration of the wall pressure and shear stress and from the Josephson-Anderson relation. (\fullgreen{}) Total drag work evaluated by surface integration of pressure and friction; (\full{}) time-averaged drag work evaluated by J-A relation; (\dashedgreen{}) pressure work; (\chaingreen{}) friction work; (\dashedblack{}) total advective vorticity flux $\int_{\Omega}\boldsymbol{u}_{\phi}\cdot\left(-\boldsymbol{u}\times\boldsymbol{\omega}\right)\mathrm{d}V$; (\chainblack{}) total viscous vorticity flux $\int_{\Omega}\boldsymbol{u}_{\phi}\cdot\left(\nu\boldsymbol{\nabla}\times\boldsymbol{\omega}\right)\mathrm{d}V$; (\dottedgray{}) oscillation of JA drag. $(b)$ Symbols represent the time averaged J-A drag evaluated over a domain of radius $r$. (\dashedblack{}) Time-averaged drag force from the summation of form and friction drag.}
    \label{fig:sphere3700_history}
\end{figure}

The contributions to the drag coefficient are presented in figure \ref{fig:sphere3700_history}.  
The form drag is significantly higher than friction drag due to the higher Reynolds number considered relative to case `SL'. In addition, small temporal variations in the form and total drag can be discerned, which are due to the unsteadiness of the flow field. 
The similarity of values of form drag and advective flux contribution, as well as friction drag and viscous flux contribution, are observed in figure \ref{fig:sphere3700_history} as expected (the mathematical relations are provided in Appendix \ref{sec: appendix}).
The advective term from the J-A relation exhibits high-frequency oscillations. The finite size of the simulation domain can only accommodate a portion of the full streamwise extent of the wake, here $15$ diameters.  Vortices that leave the domain cause oscillations in the total advective vorticity flux and the rate of drag work from the J-A relation. The gray dotted line in figure \ref{fig:sphere3700_history}(a) represents the spatial integration of the J-A relation outside the domain, which is computed by subtracting the J-A drag from the surface integral drag. The exchange of circulation between the computational domain and the outside flow causes the oscillation of J-A drag inside the computational domain. To examine the influence of the domain size on the accuracy of the J-A relation, we calculated the J-A integrals on domains $\Omega_r$ with different radii,  
\begin{equation}
    \boldsymbol{F}_r \cdot \boldsymbol{U}= \rho\int_{\Omega_r} \boldsymbol{u}_{\phi}\cdot\left(\boldsymbol{u}\times\boldsymbol{\omega}-\nu\boldsymbol{\nabla}\times\boldsymbol{\omega}\right)\mathrm{d}V. 
\end{equation}
The mean value 
\begin{equation}
    \overline{\boldsymbol{F}}_r \cdot \boldsymbol{U}= \frac{1}{T_2-T_1}\int_{T_1}^{T_2} \boldsymbol{F}_r \cdot \boldsymbol{U} \mathrm{d}t,
\end{equation}
is visualized in figure \ref{fig:sphere3700_history}(b), and converges rapidly with the increase of domain radius $r$. The results indicate that the domain size is appropriate.

\begin{figure}
    \centering
    \includegraphics[width=1.0\textwidth]{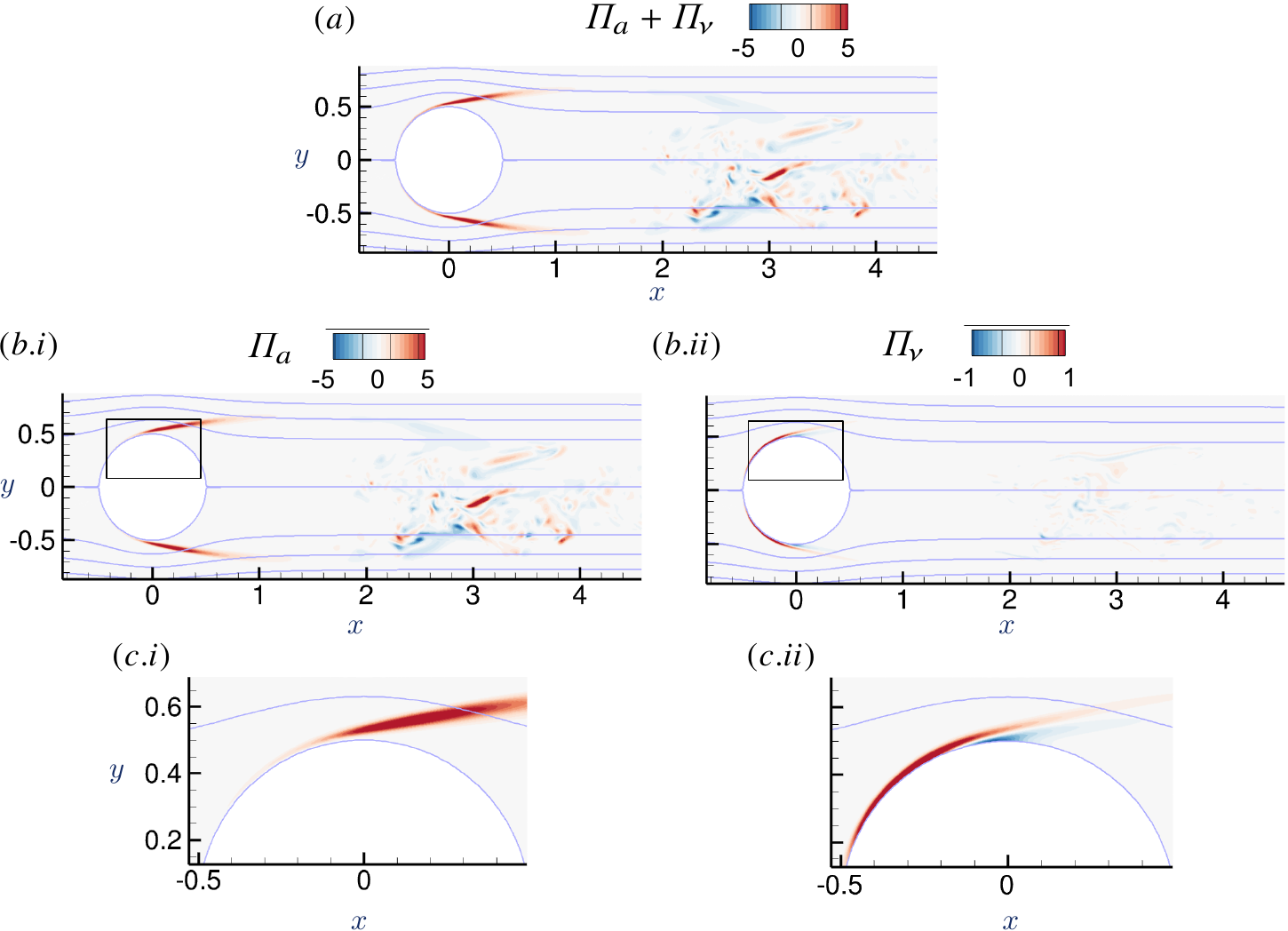}
    \caption{Two-dimensional view of the instantaneous vorticity fluxes for case `ST'. Colour contours represent $(a)$ $\Pi_a+\Pi_\nu$ , ($b.i$, $c.i$) $\Pi_a$,  and  ($b.ii$, $c.ii$) $\Pi_\nu$.  Lines correspond to the potential-flow streamlines with $\psi = \{0, -0.1, -0.2, -0.3\}$.}
    \label{fig: sphere3700 instantaneous J-A contour}
\end{figure}

The instantaneous fields of $\Pi_a$, $\Pi_{\nu}$ and their sum are visualized along with the potential-flow streamlines in figure \ref{fig: sphere3700 instantaneous J-A contour}. The advective term $\Pi_a$ dominates the total flux in most of the flow, while the viscous term $\Pi_{\nu}$ is only important near the wall. 
The advective flux $\Pi_a$ in this case exhibits similar behavior as observed in case SL in figure \ref{fig: sphere200 instantaneous J-A contour}, where the circular detached shear layer near the sphere carries vorticity outward crossing the potential streamlines and produces drag. 
Farther downstream, the shear layer breaks down carrying vorticity inward towards the centerline.  Crossing the potential streamline in this direction amounts to a force aligned with the potential flow, and thus reduces the rate of drag work. 
In this example, a direct analysis of the instantaneous vorticity transport is difficult due to the turbulence in the wake. We therefore proceed to analyze the mean vorticity transport, and exploit its symmetry.

\begin{figure}
    \centering
    \includegraphics[width=1.0\textwidth]{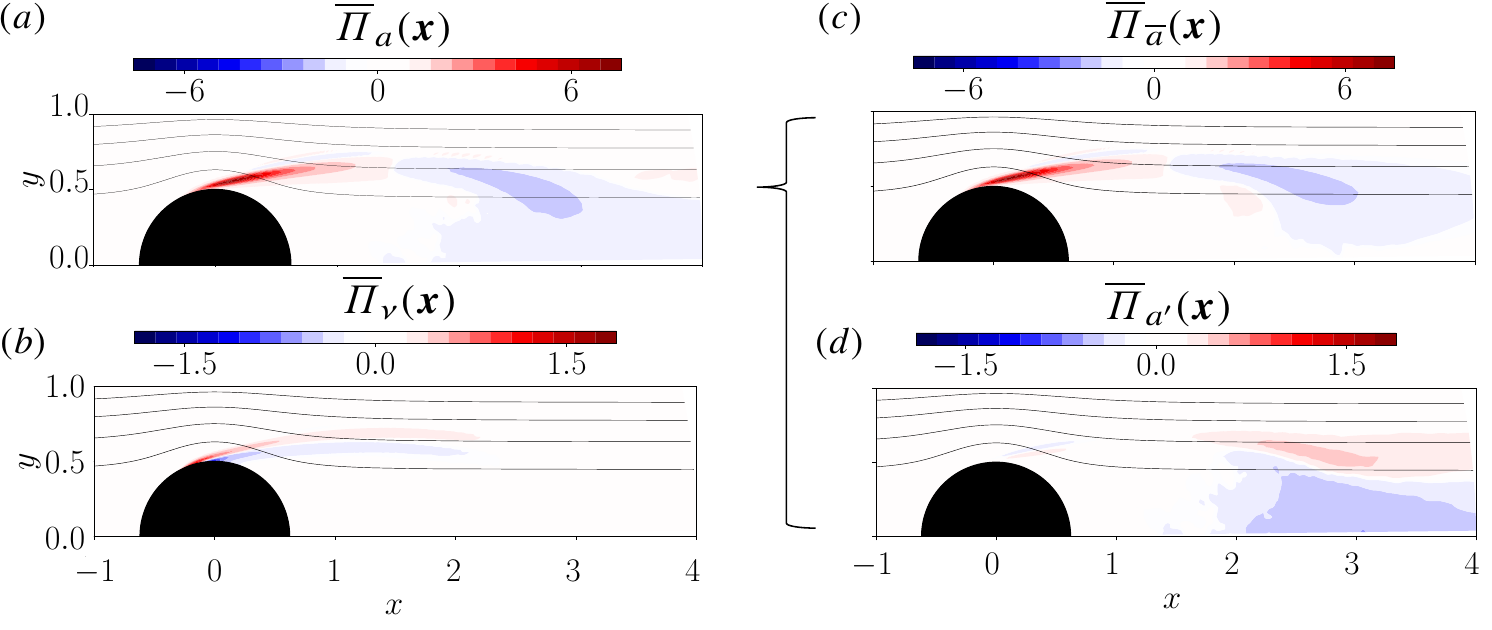}
    \caption{The mean vorticity fluxes cross the potential streamline for case ST. The solid black lines in the background represents potential function iso-surfaces $\psi=\{-0.1,-0.2,-0.3,-0.4\}$ }
    \label{fig: sphere3700 mean J-A contour}
\end{figure}

The contributions to the mean flux tensor are simplified to the following form due to the axisymmetry of the flow statistics: 
\begin{equation}
    \overline{\boldsymbol{\Sigma}}_{\overline{a}} = \begin{pmatrix}
0 & 0 & \overline{u}_x\,\overline{\omega}_{\varphi}\\
0 & 0 & \overline{u}_r\,\overline{\omega}_{\varphi}\\
-\overline{u}_x\,\overline{\omega}_{\varphi} & -\overline{u}_r\,\overline{\omega}_{\varphi} & 0\\
\end{pmatrix},\qquad
    \overline{\boldsymbol{\Sigma}}_{\nu} = \begin{pmatrix}
0 & 0 & -\nu\frac{\partial\overline{\omega}_{\varphi}}{\partial x}\\
0 & 0 & -\nu\frac{\partial\overline{\omega}_{\varphi}}{\partial r}\\
\nu\frac{\partial\overline{\omega}_{\varphi}}{\partial x} & \nu\frac{\partial\overline{\omega}_{\varphi}}{\partial r} & 0
\end{pmatrix}
\end{equation}
\begin{equation}
    \overline{\boldsymbol{\Sigma}}_{a'} = \begin{pmatrix}
0 & \overline{u_x'\omega_{r}'}-\overline{u_{r}'\omega_{x}'} & \overline{u_x'\omega_{\varphi}'}-\overline{u_{\varphi}'\omega_{x}'}\\
\overline{u_r'\omega_{x}'}-\overline{u_{x}'\omega_{r}'} & 0 & \overline{u_{r}'\omega_{\varphi}'}-\overline{u_{\varphi}'\omega_{r}'}\\
\overline{u_{\varphi}'\omega_{x}'}-\overline{u_{x}'\omega_{\varphi}'} & \overline{u_{\varphi}'\omega_{r}'}-\overline{u_{r}'\omega_{\varphi}'} & 0
\end{pmatrix},
\end{equation}
where the overbar marks mean quantities and prime denotes fluctuations. The flux of azimuthal vorticity is obtained by dotting the above flux tensors with the azimuthal unit-vector $\boldsymbol{e}_{\varphi}$: 
\begin{equation}
        \overline{\boldsymbol{\Sigma}}_{\overline{a}} \cdot \boldsymbol{e}_{\varphi} = \begin{pmatrix}
\overline{u}_x \overline{\omega}_{\varphi}\\
\overline{u}_r \overline{\omega}_{\varphi} \\
0\\
\end{pmatrix}, \qquad
    \overline{\boldsymbol{\Sigma}}_{\nu}\cdot \boldsymbol{e}_{\varphi} = \begin{pmatrix}
-\nu\frac{\partial\overline{\omega}_{\varphi}}{\partial x} \\
-\nu\frac{\partial\overline{\omega}_{\varphi}}{\partial r} \\
0\\
\end{pmatrix}, \qquad
    \overline{\boldsymbol{\Sigma}}_{a'}\cdot \boldsymbol{e}_{\varphi} = \begin{pmatrix}
\overline{u_{x}'\omega_{\varphi}'}-\overline{u_{\varphi}'\omega_{x}'} \\
\overline{u_{x}'\omega_{\varphi}'}-\overline{u_{\varphi}'\omega_{r}'} \\
0\\
\end{pmatrix}.
\end{equation}
The physical interpretation of the mean advection and diffusion fluxes of azimuthal vorticity is similar to the laminar case (\S \ref{subsection:Re200}). The turbulent flux $\overline{\boldsymbol{\Sigma}}_{a'}\cdot \boldsymbol{e}_{\varphi}$ is a combination of instantaneous advection and tilting effects. For example, the flux of azimuthal vorticity in the radial direction $\overline{u_r'\omega_{\varphi}'}-\overline{u_{\varphi}'\omega_{r}'} $ is comprised of (a) the transport of vorticity fluctuation $\omega_{\varphi}'$ by the fluctuating radial velocity $u_{r}'$ and (b) the tilting of fluctuating radial vorticity $\omega_{r}'$ into the azimuthal direction by the azimuthal fluctuation velocity $u_{\varphi}'$. 

With the detailed expressions for the vorticity fluxes, the mean J-A relation can be derived by ensemble averaging equation (\ref{eq:JA1}), which yields, 
\begin{eqnarray}
    -\overline{\boldsymbol{F}} \cdot \boldsymbol{U} &=& -\rho \int_{\Omega}\boldsymbol{u}_{\phi}\cdot\left(\overline{\boldsymbol{u}}\times\overline{\boldsymbol{\omega}}+\overline{\boldsymbol{u}'\times\boldsymbol{\omega}'}-\nu\boldsymbol{\nabla}\times\overline{\boldsymbol{\omega}}\right)\mathrm{d}V \\
        &=& -\rho \int\mathrm{d}J\int\boldsymbol{e}_{n}\cdot\left(\overline{\boldsymbol{\Sigma}}_{\overline{a}}+\overline{\boldsymbol{\Sigma}}_{a'}+\overline{\boldsymbol{\Sigma}}_{\nu}\right)\cdot\boldsymbol{e}_{\varphi} \mathrm{d}l. 
\end{eqnarray}\label{eq: mean JA}
In this equation, the rate of work done by the mean drag is related to the spatial integration of mean advection $\overline{\Pi}_{\overline{a}} = \boldsymbol{u}_{\phi}\cdot\left(\overline{\boldsymbol{u}}\times\overline{\boldsymbol{\omega}}\right)=|\boldsymbol{u}_{\phi}|\boldsymbol{e}_{n} \cdot\overline{\boldsymbol{\Sigma}}_{\overline{a}}\cdot\boldsymbol{e}_{\varphi}$,
turbulent advection $\overline{\Pi}_{a'} = \boldsymbol{u}_{\phi}\cdot\left(\overline{\boldsymbol{u}'\times\boldsymbol{\omega}'}\right)=|\boldsymbol{u}_{\phi}|\boldsymbol{e}_{n} \cdot\overline{\boldsymbol{\Sigma}}_{a'}\cdot\boldsymbol{e}_{\varphi}$, 
and diffusion $\overline{\Pi}_{\nu}= \boldsymbol{u}_{\phi}\cdot\left(-\nu\boldsymbol{\nabla}\times\overline{\boldsymbol{\omega}}\right)=|\boldsymbol{u}_{\phi}|\boldsymbol{e}_{n} \cdot\overline{\boldsymbol{\Sigma}}_\nu\cdot\boldsymbol{e}_{\varphi}$. The spatial distributions of these terms are shown in figure \ref{fig: sphere3700 mean J-A contour}. The terms $\overline{\Pi}_{\overline{a}}$, $\overline{\Pi}_{a'}$ and $\overline{\Pi}_{\nu}$ can also be interpreted as fluxes of mean azimuthal vorticity crossing iso-surfaces of the potential-flow streamfunction, and weighted by the potential flow speed. The mean term $\overline{\Pi}_{\overline{a}}$ dominates the advective flux $\overline{\Pi}_{a}=\overline{\Pi}_{\overline{a}}+\overline{\Pi}_{a'}$. Effectively, at $x<1.5$ the mean flow carries the mean azimuthal vorticity in the detached shear layer outwards, crossing the potential flow, which pushes the flow away from the ideal flow and generates drag. Further downstream, the mean flow advects the mean azimuthal vorticity inwards towards the wake centerline, which shifts the flow back towards the potential flow and corresponds to anti-drag. The turbulent flux $\overline{\Pi}_{a'}$ dominates near the wake centerline, at $x>2$. In this region, the turbulent flux accelerates the advection towards the wake centerline where the vorticity is annihilated. Note that the turbulent flux includes not only advection by the fluctuating velocity field, but also tilting of the fluctuating vorticity by the turbulent velocity.

\begin{figure}
    \centering
    \includegraphics[width=1.0\textwidth]{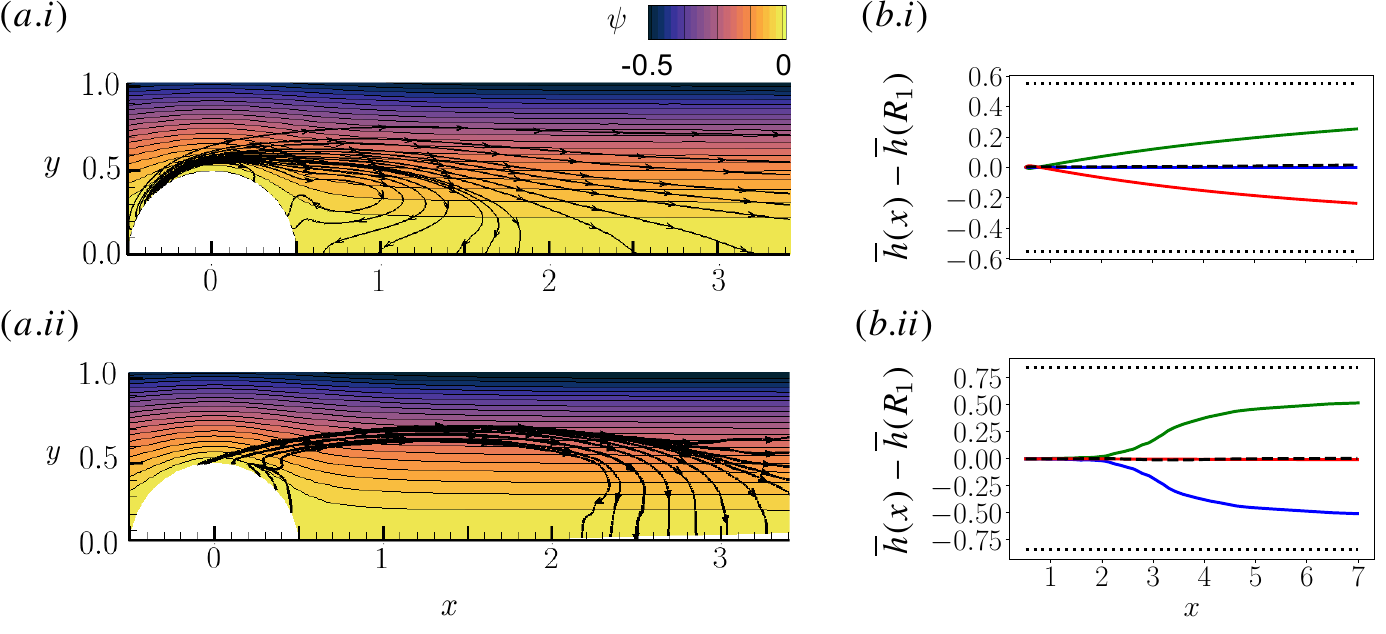}
    \caption{Balance between pressure gradient and vorticity fluxes. Panels $(a.i)$ and $(a.ii)$ show the flux lines tangent to $\left(\overline{\boldsymbol{\Sigma}}_a+\overline{\boldsymbol{\Sigma}}_\nu\right)\cdot\boldsymbol{e}_{\varphi}$ (black lines with arrow) and iso-contours of the stream-function $\psi$ (line and coloured contours) for cases SL and ST. Panels $(b.i)$ and $(b.ii)$ show the corresponding total pressure recovery, with lines representing total pressure difference (\fullgreen) $h(x)-h(R_1)$, (\fullred) $T_{\nu}$, (\full) $T_{a'}$, (\dotted) $\pm\left(h_\infty-h_{R_1}\right)$}
    \label{fig: flux_trace_presgrad}
\end{figure}

The mean momentum equation relates the annihilation rate of mean azimuthal vorticity to the total pressure gradient, according to,
\begin{equation}\label{eq: momt_wake_turbulent}
    \left(\overline{\boldsymbol{\Sigma}}_{a'}\right)_{r\varphi }+\left(\overline{\boldsymbol{\Sigma}}_\nu\right)_{r\varphi }=\overline{u_r'\omega_{\varphi}'} - \overline{u_{\varphi}'\omega_{r}'}-\nu\frac{\partial \overline{\omega}_\varphi}{\partial r}=\frac{\partial \overline{h}}{\partial x}. 
\end{equation}
Integrating equation (\ref{eq: momt_wake_turbulent}) along the wake centerline from point $x'=R_1$ to a downstream location $x'=x$ yields a relation between the total pressure recovery and the vorticity flux along this line segment, 
\begin{equation}\label{eq: momt_wake_turbulent_integrated}
   \overline{ T}_{a'} + \overline{T}_{\nu} = \int_{R_1}^{x}\left(\overline{\boldsymbol{\Sigma}}_{a'}\right)_{r\varphi }\mathrm{d}x'+\int_{R_1}^{x}\left(\overline{\boldsymbol{\Sigma}}_\nu\right)_{r\varphi}\mathrm{d}x'=\overline{h}(x)-\overline{h}(R_1). 
\end{equation}
The flux lines of azimuthal vorticity and the recovery of total pressure along the wake centerline are both visualized in figure \ref{fig: flux_trace_presgrad}, for both SL and ST cases. Overall the direction of the flux lines is similar in the laminar and turbulent cases. The azimuthal vorticity is generated at the wall, and annihilated along the wake centerline. However, the major mechanism of vorticity annihilation is different for the two cases. For case SL, the recovery of total pressure is mainly due to the viscous flux $\left(\overline{\boldsymbol{\Sigma}}_{\nu}\right)_{r\varphi}$, which is active at all $x>R_1$ and contributes a gradual recovery towards the free-stream condition. For case ST, the mean viscous flux is no-longer important. Instead, the turbulent flux balances the mean total pressure gradient. Within the recirculation bubble, $R_1<x<2$, the turbulence is nearly absent, thus the total pressure remains nearly constant. Beyond this region, the detached shear layer starts to break down. The turbulence advection and tilting contribute to the flux of azimuthal vorticity towards the wake centerline. 
This physical interpretation of the vorticity transport reinforces our understanding of the mechanism of drag, by aid of the JA relation and analysis of the Huggins flux tensor.

\section{Three-dimensional boundary-layer separation over a spheroid} \label{sec:spheroid results}

\begin{figure}
    \centering
    \includegraphics[width=0.8\textwidth]{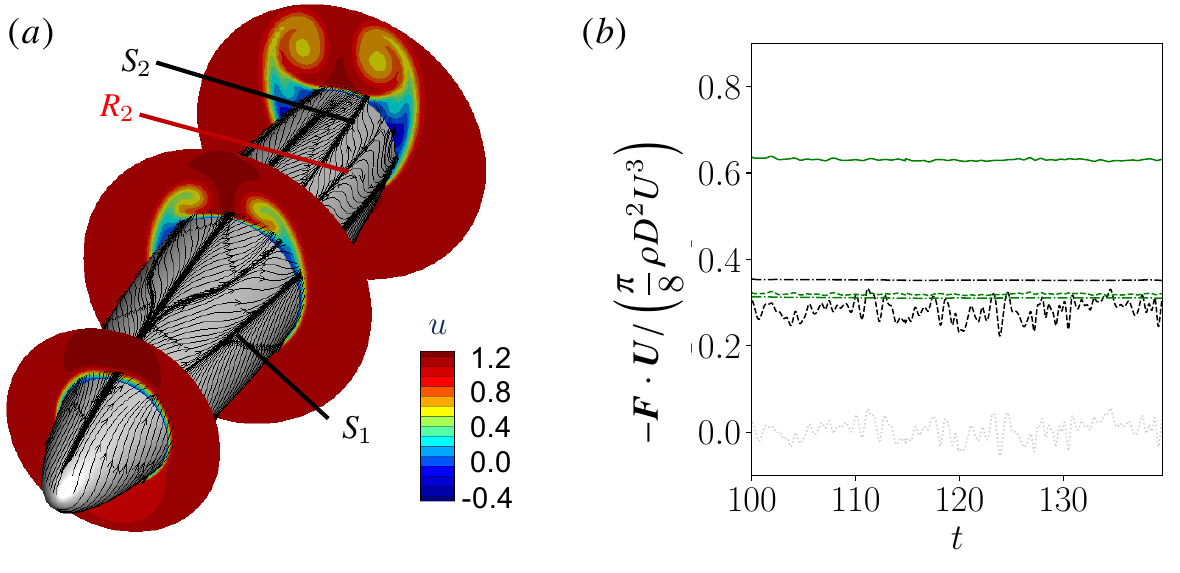}
    \caption{$(a)$ Visualization of friction lines on the surface of the spheroid and the velocity field on selected vertical planes, $x/a=\{0.16,0.5,0.84\}$. $(b)$ Time history of the drag coefficient from the surface integral and the J-A relation. (\fullgreen{}) Total drag work evaluated by surface integration of pressure and friction; (\dashedgreen{}) pressure work; (\chaingreen{}) friction work; (\dashedblack{}) total advective vorticity flux $\int_{\Omega}\Pi_a\mathrm{d}V$; (\chainblack{}) total viscous vorticity flux $\int_{\Omega}\Pi_\nu\mathrm{d}V$; (\dottedgray{}) oscillation of JA drag.}
    \label{fig:spheroid3000_history}
\end{figure}

The vorticity transport and the interpretation of drag power injection over a bluff body becomes more complex in three-dimensional separation. In this section, we consider flow over a prolate spheroid at incidence angle $\alpha = 20^{\circ}$. A visualization of the near-wall flow is shown in figure \ref{fig:spheroid3000_history}. The free-stream flow encounters the spheroid body at the stagnation point near the nose on the windward side. A thin boundary layer forms on the forebody and extends downstream. On the windward side, the attached boundary layer grows along the axial direction. On the two sides of the body, the boundary layers separate due to the combined effects of pressure gradient and curvature. The detached shear layers curve inwards towards the recirculation region and form a pair of counter-rotating vortices. These dominant vortices extend to form the turbulent far wake. The large-scale vortices induce a complex pattern of wall shear stress, visualized by the friction lines on the body surface in figure \ref{fig:spheroid3000_history}(a). A primary separation line ($S_1$) appears where the boundary layer detaches from the body. The reverse flow induced by the dominant vortices near $\varphi = 180^{\circ}$ separates again at the secondary separation line $S_2$. A reattachment line $R_2$ appears between $S_1$ and $S_2$. 
These separation and reattachment lines are well identified by the strong convergence and divergence of the wall-friction lines. We first investigate the balance given by the J-A relation and analyze the regions that contribute most to drag work, then proceed to analyze the vorticity transport and its relation to separation.

The time histories of the contributions to drag power injection were evaluated from the surface forces and from the J-A relation, and are reported in 
figure \ref{fig:spheroid3000_history}.  Although the Reynolds number is similar to turbulent flow over a sphere (case ST), the form drag here is of similar magnitude as the friction drag due to the slender shape of the body. Turning to the terms in the J-A relation, the advective contribution oscillates due to the unsteadiness of the velocity and vorticity fields, caused by the breakdown of the primary vortices.

\begin{figure}
    \centering
    \includegraphics[width=0.8\textwidth]{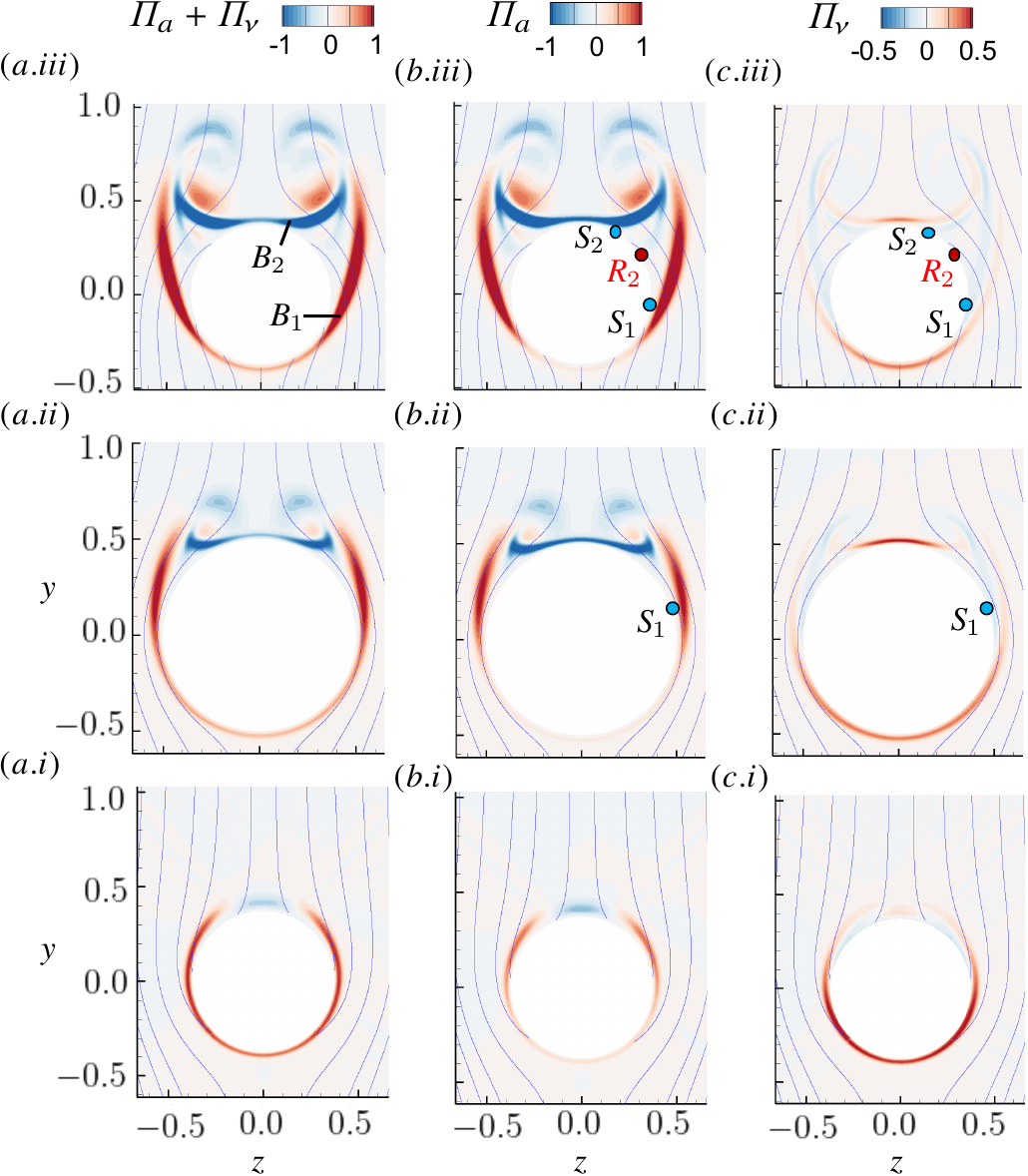}
    \caption{Contours of the instantaneous values of $(a)$ $\Pi_a+\Pi_\nu$, $(b)$ $\Pi_a$, and $(c)$ $\Pi_\nu$ for the flow over the spheroid.  Overlaid on the contours are potential-flow streamlines.  The vertical sections in the three rows are taken at $x/a=\{0.16,0.5,0.84\}$. The labels $B_1$ and $B_2$ identify the primary and secondary boundary layers, $S_1$ and $S_2$ mark the primary and secondary separations, and $R_2$ is the secondary re-attachment. The primary re-attachment is on the lee-ward plane of symmetry, and is not marked on the figure.}
    \label{fig:spheroid3000 inst JA}
\end{figure}

The instantaneous spatial contributions to the rate of drag work, $\Pi_a(\boldsymbol{x})= -\boldsymbol{u}_{\phi}\cdot\left(\boldsymbol{u}\times\boldsymbol{\omega}\right) $ and $\Pi_{\nu}(\boldsymbol{x}) = \boldsymbol{u}_{\phi}\cdot\left(\nu\boldsymbol{\nabla}\times\boldsymbol{\omega}\right)$, and their sum are visualized in figure \ref{fig:spheroid3000 inst JA} in vertical planes at the three locations marked in figure \ref{fig:spheroid3000_history}$(a)$, $x/a=\{0.16,0.5,0.84\}$. The contours are overlapped with the potential-flow streamlines. On each vertical plane, the boundary layer starts from the stagnation point at $\varphi=0$ on the wind-ward side, and separates at different azimuthal angle. The separated layer, marked $B_1$ in the figure, crosses potential streamlines carrying vorticity into the free stream. This advection of vorticity shifts the flow away from being ideal and thus contributes to drag power injection through $\Pi_a$. The secondary boundary layer, marked $B_2$ in the figure, is induced on the leeward side by the large-scale vortices, and has a favorable influence of reducing drag due to the reversed sign of $\omega_x$. Further away from the body, the pair of primary vortices advect vorticity across the potential-flow streamlines, transporting vorticity inwards at the top and outwards at the bottom, thus simultaneously contributing to and detracting from drag in these two regions, respectively.  The viscous flux $\Pi_{\nu}(\boldsymbol{x})$ is mostly concentrated inside the attached primary and secondary boundary layers, where azimuthal vorticity is generated by streamwise pressure gradient and diffuses outwards.

\begin{figure}
    \centering
    \includegraphics[width=1.0\textwidth]{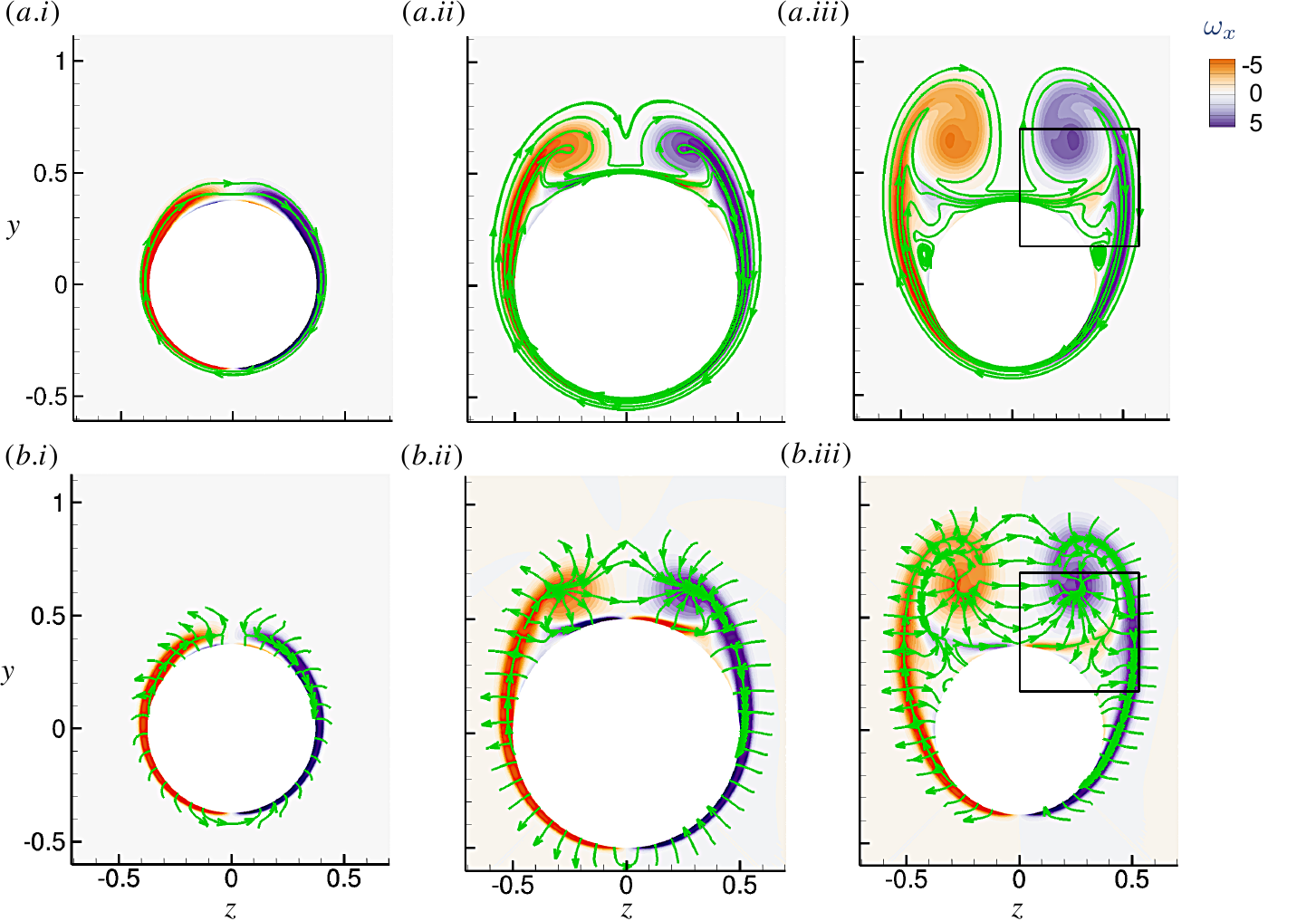}
    \caption{Contours of the streamwise vorticity, overlaid by the vorticity-flux vectors. The lines are tangent to the vector fields ($a$) due to the advective flux $\boldsymbol{\Sigma}_a\cdot\boldsymbol{e}_{x}$ and ($b$) viscous flux $\boldsymbol{\Sigma}_{\nu}\cdot\boldsymbol{e}_{x}$. (i-iii) The three panels correspond to the vertical sections at $x/a=\{0.16, 0.5, 0.84\}$. }
    \label{fig:spheroid3000 flux lines}
\end{figure}
\begin{figure}
    \centering
    \includegraphics[width=1.0\textwidth]{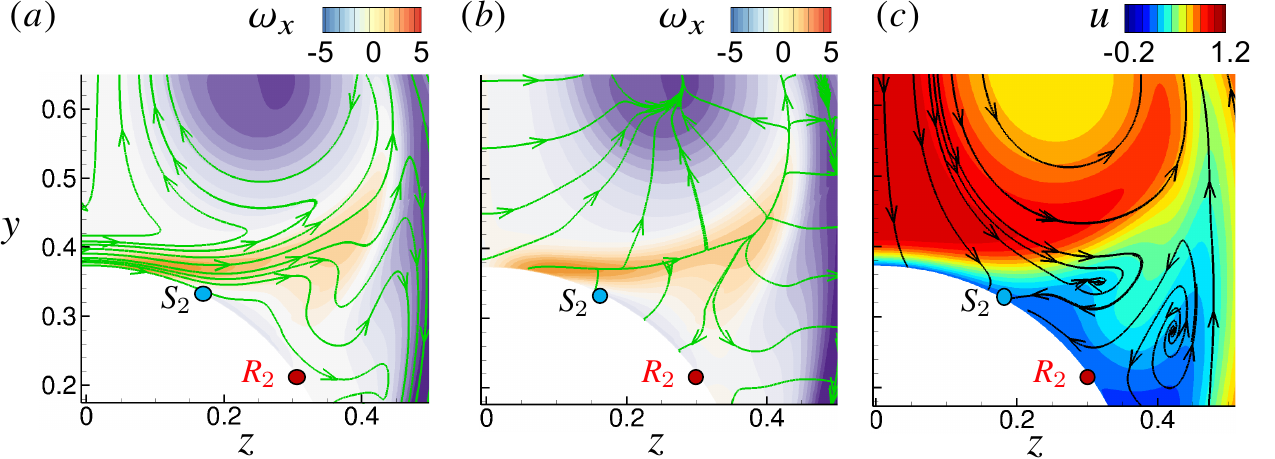}
    \caption{Visualization of streamwise vorticity and velocity contours, overlaid by the vorticity flux lines and the velocity vectors.  The shown region is the boxed area in figure \ref{fig:spheroid3000 flux lines}. The lines are tangent to $(a)$ $\boldsymbol{\Sigma}_{a}\cdot\boldsymbol{e}_{x}$, $(b)$ $\boldsymbol{\Sigma}_{\nu}\cdot\boldsymbol{e}_{x}$, and $(c)$ $\boldsymbol{u}$.}
    \label{fig:spheroid3000 VIS}
\end{figure}
\begin{figure}
    \centering
    \includegraphics[width=0.8\textwidth]{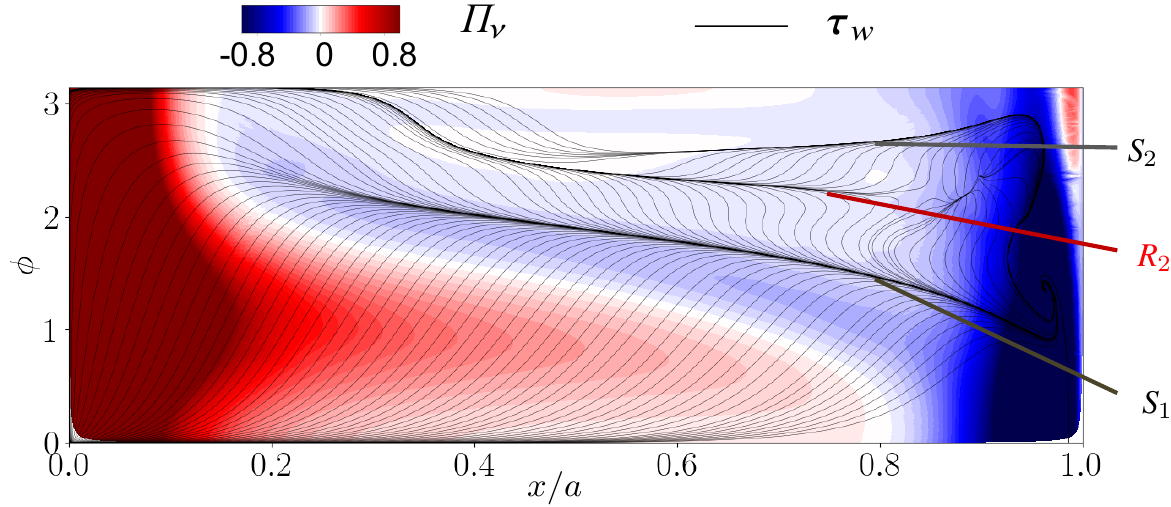}
    \caption{Visualization of wall vorticity flux and friction lines. Color: viscous flux $\Pi_{\nu}$. Lines: friction lines that are tangent to wall shear stress $\boldsymbol{\tau}_w$. }
    \label{fig:spheroid3000 wall stress line}
\end{figure}

For a detailed description of the transport of vorticity, we examine the Huggins flux tensor, specifically as it appears in the planar vorticity conservation equation (\ref{eq:planar conservation law}). We stress that this description is Eulerian, and thus focused on how the vorticity is transported locally.
We consider $\omega_{x} = \boldsymbol{\omega}\cdot \boldsymbol{e}_{x}$ since the streamwise vorticity is most relevant to the three-dimensional separation in the present flow. The vector $\boldsymbol{\Sigma}\cdot\boldsymbol{e}_{x}$ represents the two-dimensional transport of $x$-vorticity inside the vertical plane. The advection $\boldsymbol{\Sigma}_a\cdot\boldsymbol{e}_{x}$ and viscous $\boldsymbol{\Sigma}_{\nu}\cdot\boldsymbol{e}_{x}$ components are visualized in figures \ref{fig:spheroid3000 flux lines} at the same three axial locations examined in figures \ref{fig:spheroid3000_history} and \ref{fig:spheroid3000 inst JA}. Near the wall, the advection flux is parallel to the surface and the viscous flux lines start from the wall. 
The streamwise vorticity is generated at the wall at low azimuthal angles by the azimuthal pressure gradient, diffuses into the fluid. Within the detached primary boundary layer, vorticity is primarily advected.
The dominant vortices are 
regions of accumulation of streamwise vorticity in the recirculation region, as shown in the planes $x/a=\{0.5,0.84\}$ in figure \ref{fig:spheroid3000 flux lines}. A zoomed-in view in figure 
\ref{fig:spheroid3000 VIS} shows one of the two counter-rotating vortices, which induces a secondary boundary layer (and similarly for the other vortex across the vertical symmetry plane). The induced boundary layer contains $\omega_x$ of the opposite sign to the vorticity in the large-scale vortex. The secondary boundary layer detaches again from the leeward side, which is known as the secondary separation \citep{wang1990three,wetzel1996unsteady}. The streamwise vorticity $\omega_x$ is advected along this separated layer, and diffuses towards the large-scale vortex and the surface. Note that a secondary vortex appears near the end of this secondary layer, shown by the streamlines in figure \ref{fig:spheroid3000 VIS}(c).

The wall-friction lines and vorticity-flux contours are plotted in figure \ref{fig:spheroid3000 wall stress line}. The location of the separation and reattachment lines can be easily identified by the converging/diverging shear-stress lines. The viscous flux is positive at lower azimuthal locations and becomes negative upstream of the primary separation. Flow migrating from the upstream region carry excess vorticity that needs to be absorbed into the wall before separation, thus the wall vorticity flux changes sign prior to the primary separation line.

\begin{figure}
    \centering
    \includegraphics[width=1.0\textwidth]{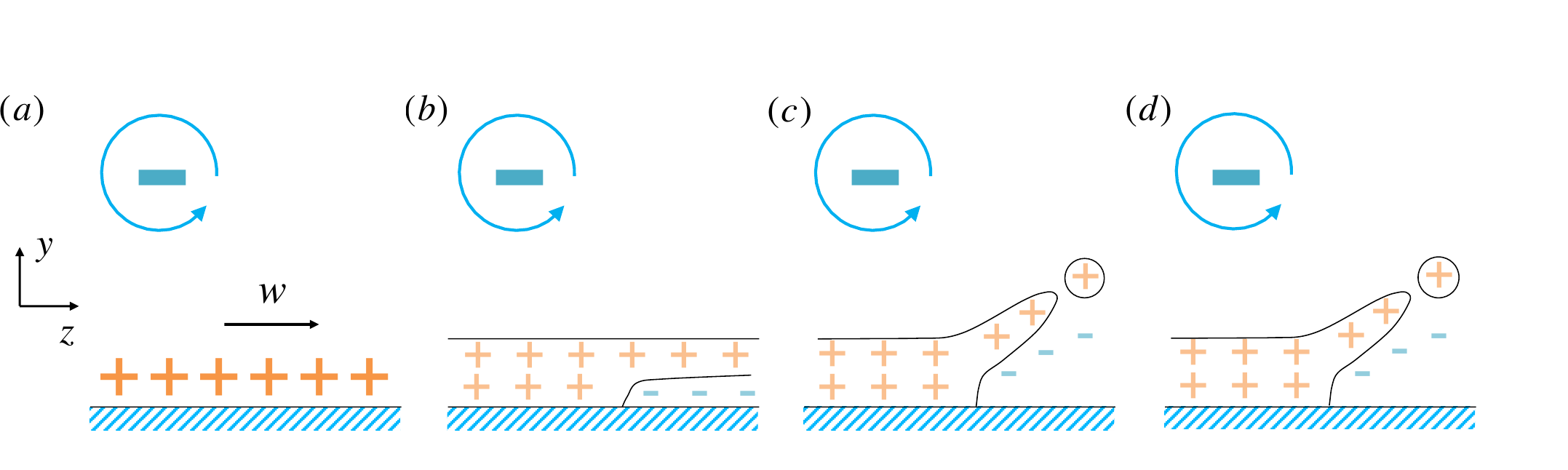}
    \caption{Schematic of vortex-induced separation, following the lecture notes  on turbulence theory by G.\,Eyink. Positive and negative symbols represent the signs of the local streamwise vorticity and color indicates the magnitude. Positive values point into the page.}
    \label{fig:vortex induced separation}
\end{figure}

In the rest of this section, we discuss the mechanism of vortex-induced secondary separation. 
The interpretation is motivated by the local flow structures and invokes ideas from two-dimensional separation. We stress, however, that the secondary separation on the spheroid is three-dimensional, and that the present discussion is intended to build intuition regarding the local flow near the secondary separation. 
Following \citet{doligalski1994vortex}, 
consider an inviscid vortex with negative circulation above a solid wall, as shown schematically in figure \ref{fig:vortex induced separation}(a). Since the wall is stationary, a vortex sheet is introduced at the surface, with strength $\gamma_w=\boldsymbol{n}\times\boldsymbol{w}(z)$ where $\boldsymbol{w}(z)$ is the fluid velocity. The velocity and pressure satisfy the inviscid balance,
\begin{equation}
    -\frac{\mathrm{d}p}{\mathrm{d}z}=w\frac{\mathrm{d}w}{\mathrm{d}z}. 
\end{equation}
Since the singular vortex sheet is proportional to the velocity magnitude, the above equation can be regarded as a balance between the destruction of the vortex sheet by pressure gradient and the advection of the sheet. If a small amount of viscosity is introduced into the flow, the singular vortex sheet becomes finite in thickness, as shown in figure \ref{fig:vortex induced separation}(b). The vorticity magnitude reduces, thus the advection of vorticity becomes smaller than its destruction by the pressure gradient. A thin layer of negative vorticity thus appears near the wall. This region of reverse flow ejects the boundary layer into the free stream, and the detached boundary layer could break up and form a secondary vortex, as shown in figures \ref{fig:vortex induced separation}(c,d). 

We return to the flow field and the vorticity flux above the spheroid at $x/a=0.83$, which are visualized in 
figure \ref{fig:spheroid3000 VIS}. The dominant vortex with negative streamwise vorticity induces the secondary boundary layer on the leeward surface with positive vorticity (panel $(b)$). This boundary layer separates due to an adverse pressure gradient imposed by the dominant vortex. Streamwise vorticity is advected along the secondary boundary layer and ejected to form a secondary vortex. Viscous diffusion of positive $\omega_x$ points from the secondary boundary layer towards the wall, where the outflux is related to the azimuthal pressure gradient. These observations are consistent with the vortex-induced separation mechanism described in figure \ref{fig:vortex induced separation}.

\section{Conclusions} \label{sec:conclusions}

In this work, we performed a numerical investigation of the vorticity dynamics and its relation with the drag force for the flow over two a bluff body. We first invoked the Josephson-Anderson relation that expresses the rate of work done by the drag force as the spatial integration of advective and viscous vorticity fluxed across the potential-flow streamlines. To understand the contribution to drag power injection by different flow regions, we investigated the Huggins vorticity flux tensor. Three numerical simulations of flow over a bluff body were performed: The flows over a sphere at $\Re = \{200, 3700\}$ and the flow over a prolate spheroid at $\Re = 3000$ and $20^{\circ}$ incidence angle. For each of these flows, the balance of the J-A relation, the  contributions to drag from the spatially distributed vorticity fluxes, and the flux vectors were computed and visualized. Our analysis addressed the most relevant features of each flow, including steady and unsteady two-dimensional separations, turbulent wake, and three-dimensional separation.

For the flow over a sphere at $\Rey=200$, the power injection of drag force calculated from the J-A relation and the integration of the surface stresses agree well. At the sphere surface, the azimuthal vorticity is diffused into the fluid interior by viscosity, which contributes to drag through the viscous term in the J-A relation. The vorticity in the boundary layer is first advected outwards crossing the potential-flow streamlines thus contributing to drag, and is then advected inwards towards the wake centerline thus contributing anti-drag. At the higher Reynolds number $\Rey=3700$, we first find that the drag from the J-A relation agrees well with the integration of the surface stresses during the impulsively starting stage. The J-A relation attributes all of the drag at $t=0^+$ to the viscous vorticity flux across the wall, while the friction and pressure drag are both nonzero at that time. The development of multiple unsteady separations on the sphere wall is accompanied with the ejection of the primary and secondary vortices into the fluid interior, which contributes to the advection part of drag power injection. During the statistically stationary stage, the turbulent transport of vorticity in the near wake becomes important. The annihilation of vorticity along the wake centerline is balanced by the gradient of the total pressure, and is dominated by the turbulent vorticity flux. 
For the flow over a prolate spheroid at $\Rey=3000$, the large-scale axial vortices induce multiple three-dimensional separation and reattachment lines over the leeward surface. The axial vorticity is generated in the windward boundary layer and advected into the vortices in the near wake. The primary and secondary separated boundary layers contribute opposite effects, namely an increase and a reduction, to the drag force.  In addition, the separation of the secondary boundary layer was interpreted in terms of the theory of vortex-induced separation.

There are several possible avenues of future research that involve the Josepshon-Anderson relation and vorticity dynamics, in the context of the flow over a bluff body. 
First, transitional boundary-layer flows can be studied \citep{Zaki2010,wang2022origin}. Across laminar-to-turbulence transition, the wall vorticity magnitude increases significantly and an up-gradient turbulent flux of spanwise vorticity appears inside the boundary layer \citep{lighthill1986informal}. Both phenomena introduce more complex mechanisms of vorticity transport in the near-wall region that are of both theoretical and practical interest. 
Second, the Josephson-Anderson relation provides a new framework to examine drag-reduction strategies, from a vorticity dynamics point of view, including riblets \citep{garcia2011drag, choi1993direct} and superhydrophobic surfaces \citep{daniello2009drag,Jelly2014} In addition, instead of attempting to minimize the full drag term, drag-reduction strategies can be sought that target the reduction of the outward flux of vorticity.  
Third, other physical effects can be incorporated into the Josephson-Anderson relation such as a spatiotemporal body force. For example, in viscoelastic flows \citep{terrapon2004simulated, li2007polymer, Hameduddin2018}, the influence of the polymer stress can be included in the momentum equation as a distributed body force. These effects can be taken into consideration in the Josephson-Anderson relation by accounting for the body forces in the derivation. Lastly, while the present study adopted an Eulerian perspective to examine the vorticity transport, future efforts should consider the Lagrangian evolution of vorticity \citep{xiang2024Origin} in these bluff-body flows.

\backsection[Acknowledgments]{Computational resources were provided by the Maryland Advanced Research Computing Center (MARCC). We acknowledge Prof.\,G.\,Eyink for valuable discussions regarding the theoretical aspects of the Josephson-Anderson relation and vortex-induced separation.} 

\backsection[Funding]
{The authors acknowledge financial support from the Office of Naval Research (N00014-20-1-2715). }

\backsection[Declaration of interests]
{The authors report no conflict of interest.}

\backsection[Author ORCIDs]
{Yifan Du,  https://orcid.org/0000-0002-2523-6795;  
Tamer A. Zaki, https://orcid.org/0000-0002-1979-7748}

\appendix

\section{Alternative forms of $\Pi_a$ and $\Pi_\nu$}
\label{sec: appendix}

In this appendix, we relate the advective flux term $\Pi_a$ to the wall pressure field, and we relate the diffusive flux term $\Pi_\nu$ to the wall shear stress. These derivations explain the observed similarities in figures \ref{fig:sphere3700 starting history} and \ref{fig:sphere3700_history}.

Starting from the expression of $\Pi_a$, and recalling the Poisson equation $\nabla^2h=-\nabla \cdot\left(\boldsymbol{u} \times \boldsymbol{\omega}\right)$ for the total pressure $h=\frac{p}{\rho}+\frac{1}{2}\vert\boldsymbol{u}\vert^2$, we can write, 
\begin{eqnarray}
\Pi_a&=&\int_\Omega\boldsymbol{u}_\phi\cdot\left(-\boldsymbol{u} \times \boldsymbol{\omega}\right)\mathrm{d}V=\cancel{\int_\Omega\nabla\cdot\left(-\phi\boldsymbol{u} \times \boldsymbol{\omega}\right)\mathrm{d}V}+\int_\Omega\phi\nabla\cdot\left(\boldsymbol{u} \times \boldsymbol{\omega}\right)\mathrm{d}V\\
&=&\int_\Omega-\phi\nabla\cdot\nabla h\mathrm{d}V=\int_\Omega\nabla\cdot\left(-\phi \nabla h\right)\mathrm{d}V+\int_\Omega\boldsymbol{u}_\phi\cdot\nabla h\mathrm{d}V\\
&=&\int_{\partial B}\left(\phi \hat{\boldsymbol{n}}\cdot\nabla h\right)\mathrm{d}S-\lim_{R\to\infty}\int_{ S_R}\left(\phi \hat{\boldsymbol{x}}\cdot\nabla h\right)\mathrm{d}S\nonumber+\int_\Omega\nabla\cdot\left(\boldsymbol{u}_\phi h\right)\mathrm{d}S\\
&=& \int_{\partial B}\left(\phi \hat{\boldsymbol{n}}\cdot\nabla h\right)\mathrm{d}S-\lim_{R\to\infty}\int_{ S_R}\hat{\boldsymbol{x}}\cdot\left(\phi \nabla h-h \nabla\phi \right)\mathrm{d}S\\
&=&\int_{\partial B}\phi \hat{\boldsymbol{n}}\cdot\nabla \left(\frac{p}{\rho}+\frac{1}{2}\vert\boldsymbol{u}\vert^2\right)\mathrm{d}S. \label{eq: advection alter}
\end{eqnarray}
The above expression establishes the connection between the advective term and the wall pressure. 

Similarly, an expression can be derived that relates the viscous term $\Pi_{\nu}$ to the wall shear stress, 
\begin{eqnarray}
    \Pi_\nu&=&\int_{\Omega} \boldsymbol{u}_\phi \cdot(\nu \boldsymbol{\nabla} \times \boldsymbol{\omega}) d V:=\int_{\Omega}\nu\partial_i\phi\epsilon_{ijk}\partial_j\omega_k\mathrm{d}V\\
    &=&\int_{\Omega}\nu\partial_j\left(\partial_i\phi\epsilon_{ijk}\omega_k\right)\mathrm{d}V-\cancel{\int_\Omega \nu\left(\partial_i\partial_j\phi\right)\epsilon_{ijk}\omega_k \mathrm{d}V}\\
    &=&:\int_{\Omega}\nu\nabla\cdot\left(-\nabla\phi\times\boldsymbol{\omega}\right)\mathrm{d}V=-\int_{\partial\Omega}\nu\boldsymbol{u}_\phi\cdot\left(\hat{\boldsymbol{n}}\times\boldsymbol{\omega}\right) \mathrm{d}V.\\
    &=&\int_{\partial\Omega}\frac{1}{\rho}\boldsymbol{u}_\phi\cdot\boldsymbol{\tau}\mathrm{d}V
\end{eqnarray}
The last expression can be interpreted as the wall shear stress exerting work along the potential flow.

\bibliographystyle{jfm}
\bibliography{main}

\end{document}